\documentclass[aps,prd,twocolumn,showpacs,notitlepage,eqsecnum,
superscriptaddress]{revtex4-1}

\usepackage[centertags]{amsmath}
\usepackage{amssymb}
\usepackage{latexsym}
\usepackage{enumerate}
\usepackage{graphicx}
\usepackage{mathrsfs}
\usepackage{hyperref}
\usepackage{stmaryrd}
\usepackage{import}
\usepackage{tensor}
\usepackage{color}
\usepackage{bm}
\usepackage{multirow}
\usepackage{breakurl}
\usepackage{float}
\usepackage{placeins}

\newcommand{\bi}{\begin{itemize}}
\newcommand{\ei}{\end{itemize}}
\newcommand{\be}{\begin{equation}}
\newcommand{\ee}{\end{equation}}

\renewcommand{\l}{\left(}
\renewcommand{\r}{\right)}
\renewcommand{\a}{\alpha}
\renewcommand{\b}{\beta}
\newcommand{\g}{\gamma}
\newcommand{\G}{\Gamma}
\renewcommand{\d}{\delta}
\newcommand{\D}{\Delta}
\newcommand{\e}{\epsilon}
\newcommand{\ve}{\varepsilon}
\newcommand{\La}{\Lambda}
\newcommand{\la}{\lambda}
\renewcommand{\O}{\Omega}
\renewcommand{\o}{\omega}

\newcommand{\q}{\quad}
\newcommand{\qq}{\qquad}

\newcommand{\vp}{\varphi}

\newcommand{\pa}{\partial}

\allowdisplaybreaks

\begin{document}

\title{Post-Newtonian expansion of the spin-precession invariant for eccentric-orbit non-spinning 
extreme-mass-ratio inspirals to 9PN and $e^{16}$}

\author{Christopher Munna}
\affiliation{MIT Kavli Institute, Massachusetts Institute of Technology, Cambridge, MA 02139, USA }
\affiliation{Department of Physics and Astronomy, University of North Carolina, Chapel Hill, North Carolina 27599, USA}
\author{Charles R. Evans}
\affiliation{Department of Physics and Astronomy, University of North Carolina, Chapel Hill, North Carolina 27599, USA}
  
\begin{abstract}
We calculate the eccentricity dependence of the high-order post-Newtonian (PN) expansion of the spin-precession 
invariant $\psi$ for eccentric-orbit extreme-mass-ratio inspirals with a Schwarzschild primary.  The series 
is calculated in first-order black hole perturbation theory through direct analytic expansion of solutions in the 
Regge-Wheeler-Zerilli formalism, using a code written in \textsc{Mathematica}.  
Modes with small values of $l$ are found via the Mano-Suzuki-Takasugi (MST) analytic function expansion formalism 
for solutions to the Regge-Wheeler equation.  Large-$l$ solutions are found by applying a PN expansion ansatz to 
the Regge-Wheeler equation.  Previous work has given $\psi$ to 9.5PN order and to order $e^2$ (i.e., the near 
circular orbit limit).  We calculate the expansion to 9PN but to $e^{16}$ in eccentricity.  
It proves possible to find a few terms that have closed-form expressions, all of which are associated with 
logarithmic terms in the PN expansion.  We also compare the numerical evaluation of our PN expansion to prior 
numerical calculations of $\psi$ in close orbits to assess its radius of convergence.  We find that the series is 
not as rapidly convergent as the one for the redshift invariant at $r \simeq 10M$ but still yielding $\sim 1\%$ 
accuracy for eccentricities $e \lesssim 0.25$.
\end{abstract}


\maketitle

\section{Introduction}
\label{sec:intro}

In a set of recent papers, we have presented high post-Newtonian (PN) order analytic expansions of black hole 
perturbation theory (BHPT) and gravitational self-force quantities at first order in the mass ratio $\ve \ll 1$ 
for extreme-mass-ratio inspiral (EMRI) binaries in bound eccentric motion about a Schwarzschild black hole.  
In each case, these results are double expansions in PN order and in powers of the eccentricity $e$.  This work 
included study in the dissipative sector of gravitational wave energy and angular momentum fluxes radiated to infinity 
\cite{Munn20,MunnEvan19a,MunnETC20,MunnEvan20a} and fluxes radiated into the horizon \cite{MunnEvan22c} and study 
in the conservative sector of the redshift invariant \cite{MunnEvan22a}.  The method involves using the 
Regge-Wheeler-Zerilli (RWZ) formalism \cite{ReggWhee57,Zeri70} and making analytic function expansions using the 
Mano-Suzuki-Takasugi (MST) formalism \cite{ManoSuzuTaka96a} and a general-$l$ ansatz to find expansions of the mode 
functions.  The metric perturbations and self-force are derived in the Regge-Wheeler (RW) gauge and mode-sum 
regularization is used.  A sampling of other applications that have used this procedure include 
\cite{BiniDamo13,BiniDamo14a,BiniDamo14b,BiniDamo14c,KavaOtteWard15,ForsEvanHopp16,HoppKavaOtte16}.  

This paper applies those techniques to another gauge-invariant quantity, the spin-precession invariant.
This invariant, $\psi$, quantifies the geodetic precession of a gyroscope attached to the smaller 
mass as it is parallel transported during its orbital motion.  The test-body limit of the geodetic precession is 
well known.  We are concerned with the first order in $\ve$ correction to $\psi$, $\D \psi$, induced by the small 
but finite mass of the secondary.  For an eccentric orbit, $\psi$ is defined as the fractional precessional angular 
advance $\Psi$, per azimuthal angular advance $\Phi$, accumulated over one radial libration.  
The calculation of $\D \psi$ bears some similarities to that of the redshift invariant, as they both depend on 
the metric perturbation at the point mass location.  As discussed in \cite{MunnEvan22a, HoppKavaOtte16}, the 
PN order of individual modes of the local metric perturbation do not increase with $l$, which means that the mode
functions and metric perturbation must be calculated for arbitrarily high $l$.  This general-$l$ complication is 
handled by utilizing a PN-expansion ansatz solution to the RW equation valid for all $l$ above the target PN order
\cite{BiniDamo13,BiniDamo14a,BiniDamo14b,KavaOtteWard15,HoppKavaOtte16,MunnEvan22a}. 

Calculating $\D \psi$ presents new challenges.  One is the need to calculate the (conservative) self-force 
itself.  (In contrast, the redshift invariant only required the metric perturbation.)   Calculation of all of the 
metric perturbation components and the components of the self-force is roughly an order of magnitude more 
computationally costly than the effort involved in finding the redshift invariant.  Furthermore, the self-force is 
gauge dependent.  Fortunately, the regularization is performed on the $l$-modes of the spin-precession invariant 
itself, extracting the gauge invariant result directly.  However, the mode-sum regularization procedure in this 
case requires two regularization parameters in order for the mode-sum to converge.

The spin-precession invariant was originally calculated for circular orbits in \cite{DolaETC14a}, both numerically 
and as a full arbitrary-mass-ratio PN expansion to 3PN absolute order.  (Note that in contrast to previous papers 
on fluxes, where we referred to relative PN orders, here we connote PN order with the power of the PN compactness 
parameter ($y$ or $1/p$) appearing in the expansion of $\D\psi$, as is conventional in papers on the spin invariant.)  
The spin-precession invariant was previously found \cite{KavaOtteWard15} to 21.5PN in the circular-orbit limit 
using BHPT analytic expansions.  In the eccentric-orbit case, results were found both numerically and as a 3PN 
expansion in \cite{AkcaDempDola17}.  Note, that the circular-orbit quantity $\D \psi^{\rm circ}$ is not the same 
as its eccentric-orbit counterpart $\D \psi^{\rm ecc}$ when the latter is taken in the limit $e \rightarrow 0$.  
The eccentric-orbit definition relies on angular changes accumulated over one radial libration.  In 
the limit as $e\rightarrow 0$, apsidal advance becomes indistinguishable from azimuthal advance but the difference 
in these definitions involves the order $\ve$ correction to the apsidal advance.  The calculation 
of the eccentric-orbit version was separately found \cite{AkcaDempDola17} to 9.5PN.  The $\mathcal{O}(e^2)$ 
correction was then computed to 3PN in \cite{AkcaDempDola17}, to 6PN in \cite{KavaETC17}, to 9PN in 
\cite{BiniDamoGera18}, and then to 9.5PN in \cite{BiniGera19b}.  The present work finds $\D\psi$ to 9PN but takes 
the eccentricity expansion to $e^{16}$, breaking away from the nearly circular orbit limit.

Conservative quantities like the spin-precession invariant supply crucial terms in effective-one-body (EOB) potentials 
\cite{BaraDamoSago10, LetiBlanWhit12, BiniDamo14c, BiniDamoGera15, HoppKavaOtte16, KavaETC17, 
BiniDamoGera18, BiniDamoGera19, BiniDamoGera20a, BiniDamoGera20b}) and also contribute directly to the EMRI 
cumulative phase at post-1 adiabatic order \cite{HindFlan08}.  A procedure is described by \cite{KavaETC17} for 
translating the expansion of $\D \psi$ to the EOB gyrogravitomagnetic potential $g_{S*}(1/r,p_r,p_\vp)$, thus 
informing the spin-orbit sector of EOB dynamics.  The spin-precession invariant expansion in this paper can be 
transcribed to EOB form to enhance further the knowledge of the spin-orbit part.

The structure of this paper is as follows: In Sec.~\ref{sec:expReview} we briefly outline (i) the setup of the 
orbital motion problem, (ii) the MST formalism for computing solutions and PN expansions of specific $l$ modes, 
(iii) the procedure for finding general-$l$ parts of the expansion, and (iv) the calculation of the (local) 
metric perturbation.  Sec.~\ref{sec:spinCalc} (i) defines the spin-precession invariant, (ii) describes the 
background tetrad and how to calculate the precession, (iii) summarizes how the first-order correction to the 
spin precession is computed with a definition that is gauge invariant, and (iv) how mode-sum regularization is 
applied to the spin invariant.  Then, the results of our calculations are presented in Sec.~\ref{sec:PNresults}, 
first as a PN expansion in the compactness parameter $1/p$ and second as an expansion in the PN parameter $y$.
Our expansions are also evaluated numerically at a pair of close orbital separations and compared to prior 
numerical calculations.  Sec.~\ref{sec:ConsConc} concludes with summary and outlook.

Throughout this paper we choose units such that $c = G = 1$, though $\eta = 1/c$ is briefly reintroduced for 
PN-expansion bookkeeping purposes.  We use metric signature $(-+++)$.  Our notation for the RWZ formalism follows 
that found in \cite{ForsEvanHopp16, MunnETC20}, which in part derives from notational changes for tensor spherical 
harmonics and perturbation amplitudes made by Martel and Poisson \cite{MartPois05}.  For the MST formalism, we largely 
make use of the discussion and notation found in the review by Sasaki and Tagoshi \cite{SasaTago03}.

\section{Formalism for black hole perturbations and post-Newtonian expansions}
\label{sec:expReview}

A pair of recent papers \cite{Munn20,MunnEvan22a} outlined our approach to calculating the first-order 
metric perturbation for eccentric-orbit non-spinning EMRIs and PN expanding regularized quantities.  The more 
recent paper used the technique to derive the high-order PN expansion of the redshift invariant.  For our
present purpose, in calculating the spin-precession invariant, and to set the notation, we briefly recite in this 
section the calculational approach.  See \cite{Munn20,MunnEvan22a} for further details.

\subsection{Bound orbits and PN compactness parameters}
\label{sec:orbits}

The secondary is treated as a point mass $\mu$ in bound geodesic orbit about a Schwarzschild black hole of mass $M$, 
with $\ve = \mu/M \ll 1$.  We use Schwarzschild coordinates $x^{\mu} = (t,r,\theta, \varphi )$ that produce the 
line element
\be
ds^2 = -f dt^2 + f^{-1} dr^2
+ r^2 \left( d\theta^2 + \sin^2\theta \, d\varphi^2 \right) ,
\ee
with $f = 1 - 2M/r$.  Restricting the motion to the equatorial plane, the four-velocity is
\be
\label{eqn:four_velocity}
u^\a(\tau) = \frac{dx_p^{\alpha}(\tau)}{d\tau} 
= \l \frac{{\mathcal{E}}}{f_{p}}, u^r, 0, \frac{{\mathcal{L}}}{r_p^2} \r ,
\ee
where $\mathcal{E}$ and $\mathcal{L}$, the specific energy and angular momentum, are constants of the motion and the 
subscript $p$ indicates evaluation along the worldline of the particle.  The orbital motion is reparameterized using 
Darwin's parameters $(\chi, p, e)$ \cite{Darw59, CutlKennPois94, BaraSago10}, connected by
\begin{align}
\label{eqn:defeandp}
{\mathcal{E}}^2 &= \frac{(p-2)^2-4e^2}{p(p-3-e^2)},
\q
{\mathcal{L}}^2 = \frac{p^2 M^2}{p-3-e^2},  
\notag 
\\
& \qq \q  r_p \l \chi \r = \frac{pM}{1+ e \cos \chi} .
\end{align}
Here $p$ is the semilatus rectum and its reciprocal $1/p$ serves as one choice for a PN compactness parameter.  
In the Darwin parameterization, one radial libration corresponds to $2 \pi$ advance in $\chi$.  Motion in the other 
three coordinates, along with proper time $\tau$, are found by integrating ordinary differential equations (ODEs) in 
$\chi$ \cite{CutlKennPois94,HoppETC15}.  Most of these equations of motion can be initially PN expanded and then 
integrated analytically order by order.  For example, the radial period is found from the following integral
\begin{align}
\label{eqn:O_r}
T_r = \int_{0}^{2 \pi}  \frac{r_p \l \chi \r^2}{M (p - 2 - 2 e \cos \chi)}
 \left[\frac{(p-2)^2 -4 e^2}{p -6 -2 e \cos \chi} \right]^{1/2}    d \chi ,
 \notag
\end{align}
and it is immediately clear how the integrand may be expanded in powers of $1/p$, resulting in a series of 
elementary trigonometric integrals.  From that expansion then follows an expansion for the radial frequency, 
$\Omega_r = 2 \pi/T_r$.  In the case of azimuthal motion, the solution for $\vp_p(\chi)$ can be obtained 
analytically prior to PN expansion
\be
\label{eqn:O_phi}
\O_\varphi = \frac{4}{T_r} \left(\frac{p}{p - 6 - 2 e}\right)^{1/2} \, 
K\left(-\frac{4 e}{p - 6 - 2 e}  \right) ,
\ee
where $K(m)$ is the complete elliptic integral of the first kind \cite{GradETC07}.  The solution can be readily PN 
expanded in $1/p$.  Once the mean angular rate is known, the alternative standard PN compactness parameter 
$y = (M \O_\vp)^{2/3}$ can be obtained in terms of $1/p$, and then inverted for $p(y)$.  For eccentric motion, 
PN expansion in $1/p$ or $y$ leads to related expansion in powers of eccentricity $e$.

\subsection{Gravitational perturbations and analytic expansion of $l$-mode solutions}
\label{sec:TDmasterEq}

On a Schwarzschild background, we can obtain metric perturbations either via the Regge-Wheeler-Zerilli (RWZ) 
\cite{ReggWhee57, Zeri70} formalism (see recent uses \cite{HoppEvan10,HoppKavaOtte16,MunnEvan22a}) or by use of 
the Bardeen-Press-Teukolsky equation and radiation gauge \cite{KavaETC17}.  In this paper, we adhere to our previous 
RWZ approach, in which the RWZ master equations have the following form in the frequency domain (FD)
\be
\label{eqn:masterInhomogFD}
\left[\frac{d^2}{dr_*^2} +\omega^2 -V_l (r) \right]
X_{lmn}(r) = Z_{lmn} (r) .
\ee
Here $r_{*} = r + 2 M \ln | r/2 M - 1 |$ is the tortoise coordinate, $\o \equiv \o_{mn} = m \O_\vp + n \O_r$ are 
discrete frequencies from the multiperiodic background geodesic motion, and the FD source term is
\begin{align}
Z_{lmn} &= \frac{1}{T_r} \int_0^{2 \pi} (G_{lm}(t) \, \delta[r - r_p(t)] \notag \\& \qq \qq \qq
+ F_{lm}(t) \, \delta'[r - r_p(t)] ) e^{i \o t} dt .
\end{align}
The source terms and potentials $V_{l}(r)$ are ($l+m$) parity-dependent.  The source terms can be found in 
\cite{HoppEvan10}.  

The homogeneous version of the master equation yields two independent (causal) solutions.  One, 
$X_{lmn}^{\rm in} = X_{lmn}^{-}$, is a downgoing wave at the future horizon, while the other, 
$X_{lmn}^{\rm up} = X_{lmn}^{+}$, is an outgoing wave at future null infinity.  The odd-parity homogeneous 
(Regge-Wheeler) equation is more readily solved.  For even parity cases of $l+m$, the Regge-Wheeler equation can 
be solved again and those solutions can be transformed to their even-parity counterparts using the the 
Detweiler-Chandrasekhar transformation \cite{Chan75,ChanDetw75,Chan83,Bern07}. 

Once the homogeneous solutions are calculated (discussed below), the inhomogeneous solutions to 
\eqref{eqn:masterInhomogFD} are found, which starts by computing the normalization coefficients
\begin{widetext}
\begin{align}
\label{eqn:ClmnIntChi}
C_{lmn}^{\pm} = \frac{1}{W_{lmn} T_r} \int_0^{2 \pi} \left(\frac{dt}{d\chi} \right) 
 \bigg[  \frac{1}{f_p}G_{lm}(\chi) X^{\mp}_{lmn}  
+\left( \frac{2 M}{r_p^2 f_p^2}  X^{\mp}_{lmn} - \frac{1}{f_p} \frac{d X^{\mp}_{lmn}}{dr} \right)  F_{lm}(\chi) \bigg] 
e^{i \o t(\chi)} d\chi ,
\end{align}
where $W_{lmn}$ is the Wronskian.  The full time-domain solutions follow from applying the method of extended 
homogeneous solutions \cite{BaraOriSago08}, using the combinations $C_{lmn}^+ X_{lmn}^+$ and $C_{lmn}^- X_{lmn}^-$ 
(see also \cite{HoppEvan10, Munn20}).

As discussed in \cite{Munn20,MunnEvan22a}, solutions for the modes of the master function (at least for small values 
of $l$) are determined using the MST formalism \cite{ManoSuzuTaka96a}, with an expansion in analytic functions. 
The odd-parity MST solution for $X_{lmn}^+$ (up to arbitrary normalization) is
\begin{align}
\label{eqn:XupMST}
X^{+}_{lmn} &= e^{iz} z^{\nu+1} \left(1- \frac{\e}{z}\right)^{-i \e} \sum_{j=-\infty}^{\infty} a_j (-2 i z)^{j} 
\frac{ \G(j + \nu + 1 - i \e) \G(j + \nu - 1 - i \e) }{\G(j + \nu + 3 + i \e) \G(j + \nu + 1 + i \e) } \times \notag \\
& \hspace{27em} U(j + \nu + 1 - i \e, 2 j + 2 \nu + 2, -2 i z) .
\end{align}
Here, $\nu$ is the renormalized angular momentum, defined to make the double-sided summation converge, and $U$ is 
the irregular confluent hypergeometric function.  Other quantities are $\e = 2 M \o \eta^3$, $z = r \o \eta$, 
with $\eta = 1/c$ being a reintroduced PN parameter.  To obtain a solution, $\nu$ and $a_j$ are ascertained through 
a continued fraction calculation \cite{ManoSuzuTaka96a,SasaTago03}, which in our application also then leads to 
series in $\e$ for both.  PN expansions of the other terms in \eqref{eqn:XupMST} then follow, with the result 
expressible in series in both $z$ and $\e$.

The downgoing (or in) solutions $X_{lmn}^-$ have similar function expansion
\begin{align}
X^{-}_{lmn} &= e^{-iz} \left(\frac{z}{\e} - 1\right)^{-i \e} \left(\frac{\e}{z}\right)^{i \e + 1}  
\sum_{j= -\infty}^{\infty} a_j \frac{\G(j + \nu - 1 - i\e) \G(-j - \nu - 2 - i \e)}{\G(1 - 2i\e)} \times \notag \\
& \hspace{20em} {}_2F_1(j + \nu - 1 - i\e, -j - \nu - 2 - i\e; 1 - 2 i \e; 1 - z/\e) ,
\label{eqn:XinMST}
\end{align}
with $\nu$ and $a_j$ here being identical to those in \eqref{eqn:XupMST} (up to overall normalization of the latter).  
The process of expanding these homogeneous solutions by collecting on powers of $\eta$ is fully described in 
\cite{Munn20}, based on the methods presented in \cite{KavaOtteWard15}.  As described in \cite{MunnEvan22a,
KavaOtteWard15}, $z$-independent factors are removed from these solutions to reduce their complexity, since such 
factors eventually cancel through appearance in the Wronskian.

As discussed in \cite{BiniDamo13,KavaOtteWard15,Munn20,MunnEvan22a}, in conservative sector calculations 
mode-sum regularization requires summing perturbations over all $l$.  This necessitates an alternative approach of
directly PN-expanding the homogeneous version of the master equation \eqref{eqn:masterInhomogFD} for general $l$.
As shown in \cite{BiniDamo13, KavaOtteWard15}, the PN-expansion ansatz for solving the RW equation is
\begin{align}
\label{eqn:RWZans}
X_{lmn}^+ &= (z)^{-\nu} (1 + A_2 \eta^2 + A_4 \eta^4 + \cdots + A_{2l} \eta^{2l} + \mathcal{O}(\eta^{2l+1}) ) ,
\notag \\
X_{lmn}^- &= \left(\frac{\e}{z} \right)^{-\nu - 1} (1 + B_2 \eta^2 + B_4 \eta^4 + \cdots + B_{2l} \eta^{2l} 
+ \mathcal{O}(\eta^{2l+1}) ),
\end{align}
where the $A_i$ and $B_i$ are functions of $z, \e, l$.  The ansatz breaks down at PN orders at and above 
$\mathcal{O}(\eta^{2l})$.  If a target PN order $P$ is set, the ansatz will be useless for $l \le P$.  For those
finite number of modes, the MST formalism is used instead.  Once $\nu$ is found by PN-expanding the continued 
fraction calculation, the homogeneous RW equation becomes
\be
\left[ \left(1 - \frac{\e}{z} \right) \frac{\pa}{\pa z} 
\left(\left(1 - \frac{\e}{z} \right) \frac{\pa}{\pa z} \right) + \eta^2 + 
\left(1 - \frac{\e}{z} \right) 
\left(\frac{l (l + 1)}{z^2} - \frac{3 \e}{z^3} \right)\eta^2 \right] X^{\pm}_{lmn} = 0 .
\ee
The ODE is then solved order by order.  Even-parity homogeneous solutions are again found using the 
Detweiler-Chandrasekhar transformation.  

As previously noted \cite{MunnEvan22a}, the expansion of the even-parity normalization integral is the bottleneck 
in the calculation, requiring for example $\sim$7 days and 20GB of memory on the UNC supercomputing cluster 
Longleaf to reach 10PN and $e^{20}$ relative order.  Furthermore, two relative PN orders and three orders in $e$ 
are lost in constructing and regularizing the spin-precession invariant.  Thus, our expansion is restricted to 9PN 
(8PN relative order) and $e^{16}$.  

\subsection{Metric perturbation $l$-modes and non-radiative modes}
\label{sec:metPerExps}

Since we use RWZ gauge, calculation of the $l$-modes of the metric perturbation (locally) follows the procedure 
discussed in Section II of \cite{MunnEvan22a} (see also earlier work \cite{HoppKavaOtte16,HoppEvan10}).  Briefly, 
the metric components as functions of $\chi$ are
\begin{align}
\label{eqn:MPchi}
p_{rr}^{l}(\chi) &= 
\left(\frac{d \chi}{d r}\right) \sum_{mn} \frac{Y_{lm}(\pi/2,0)}{f} C^{\pm}_{lmn} e^{i m \vp -i \o t} 
\Bigg\{\left[ \left(\frac{d r}{d \chi}\right)\frac{\La(\la + 1)}{f r} - \left(\frac{d r}{d \chi}\right) 
\frac{\La}{f} A(\chi) + r \left(\frac{d A(\chi)}{d \chi}\right) \right] X^{\pm}_{lmn}(\chi) 
\notag \\ &
+ (r A(\chi) - \La ) \left(\frac{d X^{\pm}_{lmn}(\chi)}{d \chi} \right) + r \frac{d}{d \chi} 
\left[ f \left(\frac{d \chi}{d r}\right) \left(\frac{d X^{\pm}_{lmn}(\chi)}{d \chi} \right) \right] \Bigg\}, 
\notag \\
p_{tr}^{l}(\chi)  &=  \sum_{mn} Y_{lm}(\pi/2,0) C^{\pm}_{lmn} e^{i m \vp -i \o t} (-i \o) 
\left[ r \left(\frac{d \chi}{d r}\right)  \left(\frac{d X^{\pm}_{lmn}(\chi)}{d \chi} \right) 
+ r B(\chi) X^{\pm}_{lmn} (\chi)  \right] , 
\notag \\
p_{tt}^{l}(\chi) &= f^2 p_{rr}^{l,\pm}, 
\notag \\
p_{AB}^{l}(\chi) &= r^2 \O_{AB}  \sum_{mn} Y_{lm}(\pi/2,0)  C^{\pm}_{lmn} e^{i m \vp -i \o t}
\left[ f \left(\frac{d \chi}{d r}\right) \pa_{\chi} X_{lmn}^{e,\pm} + A(\chi) X_{lmn}^{e, \pm} \right],  
\notag \\ 
p_{tB}^{l}(\chi) &=   \left( \frac{f}{2} \right) \sum_{mn} X^{lm}_B(\pi/2,0) C^{\pm}_{lmn} e^{i m \vp -i \o t}
\left(\frac{d \chi}{d r}\right)   \frac{d}{d \chi} (r X^{\pm}_{lmn}) ,  
\notag \\ 
p_{rB}^{l}(\chi) &=  \left( \frac{r}{2f} \right) 
\sum_{mn} X^{lm}_B(\pi/2,0)  C^{\pm}_{lmn} e^{i m \vp -i \o t}  (-i \o) X^{\pm}_{lmn} ,
\end{align}
with
\end{widetext}
\begin{align}
\la &= \frac{1}{2}(l+2)(l-1), \qq  \qq    \La = \la + \frac{3 M}{r},    
\notag \\
A(r) &= \frac{1}{r \La} \left[ \la(\la+1) + \frac{3 M}{r} \left( \la + \frac{2 M}{r} \right) \right],
\notag \\ 
B(r) &= \frac{1}{r f \La} \left[ \la \left(1 - \frac{ 3M}{r} \right) - \frac{3 M^2}{r^2} \right] .  
\end{align}
The up ($+$) or in ($-$) mode functions are used depending upon which side in $r$ of the point mass the evaluation 
is taken.  The perturbation $l$-modes (once sums over $m$ are made) are continuous, but $C^0$, at the particle 
location.  Sums over $m$ involve application of the addition theorem for spherical harmonics.  In \cite{MunnEvan22a} 
we discuss the most efficient way of calculating PN expansions of the resulting sums over $m$ (see Section IID of 
that paper).  The spin-precession invariant is calculated from the local self-force, which involves the metric 
perturbation and its derivatives.  Expressions for the derivatives can be easily derived from \eqref{eqn:MPchi}.  
The $l$-modes of the metric perturbation derivatives are discontinuous across the particle location, a fact that 
is important in the regularization of the spin-precession invariant.  

Another aspect of computing the self-force is that, in applying a derivative with respect to $\chi$, an expansion 
over powers of eccentricity will lose an order in $e$.  Moreover, the eccentricity expansions lose a total 
of 3 orders in $e$ in moving from the self-force to the spin invariant.  The computation of high-order series 
in $e$ is the most consuming part of the construction of the metric perturbations, particularly initially in the case 
of general $l$.  However, added investigation showed that this difficulty can be avoided in the latter case by 
observing that the general-$l$ metric perturbations yield finite polynomials in $e$ on an individual PN-order basis.  
These polynomials increase in degree linearly with PN order.  Once this pattern was recognized, PN terms in the 
general-$l$ expansion could be determined in closed form with only a low-order (imbedded) polynomial in eccentricity.  
The computational bottleneck was then transferred to the specific-$l$ part of the calculation, which does not simplify 
in the same fashion.  The resulting change in the technique allowed the spin-precession invariant to be computed to 
much higher order in eccentricity at lower PN orders.  For example, it allowed us to determine the 4PN function 
$\D \psi_4$ to $e^{30}$.

Finally, to complete the metric perturbation and self-force calculation, the radiative modes must be augmented to 
include the nonradiative $l=0$ and $l=1$ modes, originally found by Zerilli \cite{Zeri70} but gauge transformed 
\cite{SagoBaraDetw08} to maintain asymptotic flatness.  We listed those modes in our previous paper 
\cite{MunnEvan22a}, and they are described more fully in \cite{HoppKavaOtte16} and in earlier papers cited therein.  

\section{Procedure for calculating the spin-precession invariant}
\label{sec:spinCalc}

\subsection{Overview}

The smaller body is assumed to be endowed with a spin $s_{\a}$, which undergoes precession during the 
orbital motion about the heavier mass.  The spin is parallel transported $D s_{\a}/d\tau = 0$ along the geodesic with 
its tangent vector $u^{\a}$, and the spin maintains its orthogonality $s_{\a} u^{\a} = 0$ and the constancy of its 
norm $s_{\a} s^{\a}$.  This spin-orbit, or geodetic, precession has a nonzero rate of advance in the test-body 
limit ($\mu = 0$) and a self-force correction at first order in the mass ratio $\ve$ and beyond.  We seek to
calculate the first-order correction to the precession in bound eccentric orbits about a nonspinning (Schwarzschild) 
primary (thus eliminating consideration of Lense-Thirring precession).  Our presentation follows that of 
\cite{DolaETC14a, AkcaDempDola17, KavaETC17}, which we summarize in this section.

The spin-precession invariant for eccentric orbits is a generalization given by \cite{AkcaDempDola17} of the 
definition used for circular orbits \cite{DolaETC14a}.  An invariant $\psi$ is defined via a ratio that involves the 
accumulated azimuthal phase $\Phi$ and the accumulated precession of the spin vector $\Psi$ over one radial libration 
period $T_r$.  Explicitly, the quantity is given by
\be
\psi = 1 - \frac{\Psi}{\Phi} .
\ee
This scalar is a function of the mass ratio $\ve$ (i.e., subject to self-force correction $\D\psi$) and orbital 
parameters.  The latter are best chosen as the observable frequencies $\O_r$ and $\O_\vp$, lending the definition 
of $\psi$ a gauge invariant character.  This procedure is directly analogous to that in constructing the redshift 
invariant \cite{MunnEvan22a}, where the frequencies were held fixed through first order in the mass ratio.  Since 
$\Phi = \O_\vp T_r$ will itself be fixed, only the self-force correction $\D\Psi$ need be computed. 

The invariant $\psi$ encodes a portion of the first-order conservative dynamics, giving it relevance to the creation 
of waveform templates for LISA.  How $\psi$ can be transcribed to the EOB gyrogravitomagnetic ratio quantity 
$g_{S*}(1/r, p_r, p_\vp)$, which partially characterizes the spin-orbit sector of the EOB Hamiltonian, was mapped 
out in \cite{KavaETC17}.  The expansion of $\psi$ helps describe the case where the smaller body is (weakly) spinning.

\subsection{Spin precession and background reference frame}

The behavior of $\Psi$ is found by computing the parallel transport of $s_\a$ along a geodesic of the perturbed 
(regularized) metric.  To facilitate the calculation, a reference frame tetrad $e^{\a}_a$ is introduced (here $\a$ 
is the spacetime index and $a$ indicates the frame element).  Then, the spin components in the frame precess 
\cite{AkcaDempDola17} according to
\begin{align}
\frac{d \textbf{s}}{d \tau} &= \textbf{w} \times \textbf{s} ,  \notag \\
(\textbf{s})_i &= e^\a_i s_\a ,  \notag \\
 (\textbf{w})_i &= - \frac{1}{2} \e_{ijk} w^{jk} ,  \notag \\ 
w_{ij} &= -g_{\a \b} e_i^\a \frac{D e_j^\b}{d \tau} .
\end{align}
The precessional angular velocity components use the base symbol $w$ instead of $\o$ to avoid confusion with 
the discrete frequency spectrum $\o_{mn}$ of the gravitational perturbations.  The angular velocity is subject to 
the choice made for the reference frame.

A suitable frame in the background ($\mu \rightarrow 0$ limit) was given by Marck \cite{Marc83} (see also 
\cite{AkcaDempDola17,KavaETC17}, as well as \cite{ChenEvan13} for a different application), aligned with one leg 
perpendicular to the orbital (equatorial) plane and another directed along the line with the primary 
\begin{align}
e_0^\mu &= u^\mu = \left(\frac{\mathcal{E}}{f}, u^r, 0, \frac{\mathcal{L}}{r^2} \right),  \notag \\
e_1^\mu &= \frac{1}{f \sqrt{1 + \mathcal{L}^2/r^2}} (u^r, f \mathcal{E}, 0, 0),  \notag \\
e_2^\mu &= (0,0, 1/r, 0),   \notag \\
e_3^\mu &= \frac{1}{r \sqrt{1 + \mathcal{L}^2/r^2}} 
\left(\frac{\mathcal{E}\mathcal{L}}{f}, \mathcal{L} u^r, 0, 1 + \frac{ \mathcal{L}^2}{r^2} \right) .
\end{align}
This polar alignment can be maintained when $\mu \ne 0$.  The frame is non-inertial and a gyroscope will appear 
to precess with frequency $\dot\Psi = w_{13}$.  (The precession can also be found by making a $U(1)$ 
transformation \cite{ChenEvan13} in the equatorial plane to an inertial frame.)  At lowest order the (background) 
geodetic angular rate is 
\begin{align}
w^{(0)}_{13} = \frac{\mathcal{E} \mathcal{L}}{r^2 + \mathcal{L}^2} .
\end{align}

\subsection{Spin precession and self-force in the perturbed frame}
\label{sec:self-force}

The accumulated phase of the spin precession is then
\be
\Psi = \oint w_{13}(\tau) d \tau = \Psi_0 + \D \Psi ,
\ee
with the integral taken over one period of radial motion.  The last part of the expression splits $\Psi$ into 0th 
and 1st order components, with $\D \Psi$ being the first-order (conservative) correction we seek.  Here, 
$\D\Psi$ refers to the perturbation measured when the orbital frequencies have been held constant.

The procedure to compute $\D \Psi$ is established in \cite{AkcaDempDola17,KavaETC17,BiniDamoGera18}.  Assuming 
that $\d \Psi$ is a correction calculated without holding the orbital frequencies fixed, $\D \Psi$ is recovered 
by subtracting off perturbations in the frequencies
\be
\D \Psi = \d \Psi - \frac{\pa \Psi_0}{\pa \O_r} \d \O_r - \frac{\pa \Psi_0}{\pa \O_\vp} \d \O_\vp .
\ee
An alternative calculation uses first-order changes in $T_r$ and $\Phi$ \cite{AkcaDempDola17}
\be
\D \Psi = \d \Psi - \frac{\pa \Psi_0}{\pa T_r} \d T_r  -  \frac{\pa \Psi_0}{\pa \Phi} \d \Phi ,
\ee
which we found more computationally convenient.  Note that henceforth, except for our use of $\Psi_0$ and 
$w^{(0)}_{13}$, lowest-order quantities will simply be denoted by their plain base symbol, while most first-order 
corrections will explicitly carry a $\d$ (perturbed frequencies) or $\D$ (fixed frequencies) prefix.  The exception
is the metric perturbation $p_{\mu\nu}$, where the notation already indicates a first-order quantity.

At first order, the integral for $\Psi$ experiences a change due to corrections in both $w_{13}$ and $\tau$.  The
result is \cite{BaraSago11, AkcaDempDola17}
\begin{align}
\d \Psi = \int_0^{2 \pi} \left( \frac{ \d w_{13}}{w_{13} } 
- \frac{ \d u^r}{u^r} \right) w_{13} \, \frac{d \tau}{d \chi} \, d \chi  .
\end{align}
The correction to $w_{13}$ can be found \cite{AkcaDempDola17,KavaETC17} via expansion of its definition 
\begin{align}
\d w_{13} = \frac{1}{2} w_{13} \, p_{\mu \nu} e_0^\mu e_0^\nu &+ \d \G_{[31]0} \\ \notag
&+ (c_{01} e_1^\mu + c_{03} e_3^\mu)  \,\,  e_{\nu [3} \nabla_{\mu} e^\nu_{1]} .
\end{align}
In deriving the expression above, a total derivative ($d/d\tau$) term has been neglected.  The first term is 
proportional to the $p_{00}$ tetrad projection of the metric perturbation (not $p_{tt}$) and the second term is the 
tetrad projection of the correction to the affine connection 
\be
\d \G_{[31]0} = (\d \G_{\mu \nu \b} - p_{\mu \la} \G^{\la}_{\nu \b} ) e^\mu_{[3}e^\nu_{1]} u^\b  .
\ee
The third term involves coefficients $c_{01}$ and $c_{03}$ that come from the variation of the tetrad
\begin{align}
c_{01} &= \frac{1}{f \sqrt{1 + \mathcal{L}^2/r^2}} (\mathcal{E} \d u^r_{BS} - u^r \d \mathcal{E}_{BS}),  \\
c_{03} &= \frac{\d \mathcal{L}_{BS}}{r \sqrt{1 + \mathcal{L}^2/r^2}} .
\end{align}

Within these latter coefficients are terms, $\d \mathcal{E}_{BS}$ and $\d \mathcal{L}_{BS}$, that are 
$\chi$-dependent conservative corrections to the (specific) energy and angular momentum defined by Barack and 
Sago \cite{BaraSago11}
\begin{align}
\d \mathcal{E}_{BS}(\chi) &= \d \mathcal{E}_{BS}(0) -\int_0^{\chi} F_t^{\rm cons} \frac{d \tau}{d \chi'} \, d \chi', 
\\
\d \mathcal{L}_{BS}(\chi) &= \d \mathcal{L}_{BS}(0) +\int_0^{\chi} F_\vp^{\rm cons} \frac{d \tau}{d \chi'} \, d \chi'.
\end{align}
The first term (integration constant) in each of these equations is the shift that occurs at periastron.  These 
are explicitly shown \cite{BaraSago11} to be
\begin{widetext}
\begin{align}
\d \mathcal{E}_{BS}(0) &= \frac{(1+e)^2 (p - 2 - 2e)}{4e (p - 3 - e^2)} 
\left[ \frac{(1-e)^2 (p - 2 + 2e) }{M p^{3/2} \sqrt{(p - 2)^2 - 4 e^2}} 
\int_0^\pi F_\vp^{\rm cons} \frac{d \tau}{d \chi} d\chi
+ \int_0^\pi F_t^{\rm cons} \frac{d \tau}{d \chi}  d\chi  \right] ,  
\\
\d \mathcal{L}_{BS}(0) &= \frac{ M p^{3/2} \sqrt{(p - 2)^2 - 4 e^2} }{4e (p - 3 - e^2)} 
\left[ \frac{(1-e)^2 (p - 2 + 2e) }{M p^{3/2} \sqrt{(p - 2)^2 - 4 e^2}} 
\int_0^\pi F_\vp^{\rm cons} \frac{d \tau}{d \chi} d\chi 
+ \int_0^\pi F_t^{\rm cons} \frac{d \tau}{d \chi}  d\chi \right].
\end{align}
\end{widetext}
Lastly, $\d u^r_{BS}$ is the first-order correction to the radial velocity.  It can be derived from the 
normalization of the four-velocity condition (in the background spacetime \cite{BaraSago11}), which leads to
$\mathcal{E} \d \mathcal{E}_{BS} - u^r \d u^r_{BS} - r^{-2} f \mathcal{L} \d \mathcal{L}_{BS} = 0$ and from there to
\be
\d u^r_{BS} = \frac{\mathcal{E}}{u^r} \d \mathcal{E}_{BS} - \frac{f}{r^2 u^r} \mathcal{L} \d \mathcal{L}_{BS} .
\ee

As usual, the conservative part of the self-force, $F_\mu^{\rm cons}$, is given \cite{BaraSago11} by the symmetric 
combination 
\be
F_\mu^{\rm cons} = \frac{1}{2} \left( F_\mu(\chi) + \e^{(\mu)} F_\mu(-\chi) \right) , 
\ee
where $\e^{(\mu)} = (-1, 1, 1, -1)$.  Using the retarded self-force in the expression above yields a singular 
result at the particle location.  Instead, the regularized self-force can be found using the regular metric 
perturbation $p^R_{\mu\nu}$ in 
\be
F^{\a}_R = \frac{1}{2} (g^{\a \d} u^\b - 2 g^{\a \b}u^{\d} - u^\a u^\b u^\d ) u^\g \nabla_\d p^R_{\b \g} .
\ee
However, since we are concerned with computing a single scalar invariant, it is simpler to work instead with the full, 
unregularized self-force, decomposed into $l$-modes.  This leads to an $l$-mode decomposition of the 
(unregularized) spin invariant correction, with modes $\D \psi^l$.  We then apply mode-sum regularization directly 
to the spin-precession invariant. 

Hence, with the $l$-modes of the retarded metric perturbation and self-force available, it is straightforward to 
evaluate $\mathcal{E}_{BS}^l$, $\mathcal{L}_{BS}^l$, and $\G_{[31]0}^l$.  The rest of the calculation for 
$\d \Psi^l$ is then condensed \cite{BiniDamoGera18} to the following integral
\begin{widetext}
\begin{align}
\d \Psi^l = \int_0^{2 \pi} \left[ \d \G_{[31]0}^l 
+ \left(\frac{(r^2 + \mathcal{L}^2) (3M - r) \mathcal{L}}{r^5 (u^r)^2} 
- \frac{\mathcal{L}}{r^2} \right) \d \mathcal{E}_{BS}^l + \left(\frac{\mathcal{E}}{r^2}  - 
\frac{\mathcal{E} \mathcal{L}^2 (3M - r) }{r^5 (u^r)^2} \right) \d \mathcal{L}_{BS}^l  \right] 
\frac{d \tau}{d \chi} d \chi .
\end{align}
\end{widetext}
Then, $\D \Psi^l$ can be determined from $\d \Psi^l$ by removal of the frequency corrections, which involves use of
the following formulas \cite{AkcaDempDola17, BiniDamoGera18}
\begin{align}
\d T_r^l &= \int_0^{2 \pi} \left( \frac{\d \mathcal{E}_{BS}^l}{\mathcal{E}}- \frac{\d (u^r_{BS})^l}{u^r} \right) 
\frac{\mathcal{E}}{f} \frac{d\tau}{d\chi} d \chi   ,    \\
\d \Phi^l &= \int_0^{2 \pi} \left( \frac{\d \mathcal{L}_{BS}^l}{\mathcal{L}} - \frac{\d (u^r_{BS})^l}{u^r} \right) 
\frac{\mathcal{L}}{r^2} \frac{d\tau}{d\chi} d \chi  .
\end{align}

\subsection{PN and eccentricity expansion issues}

Even though we have not belabored the process, each step in this procedure involves calculating an analytic 
PN expansion using \textsc{Mathematica}.  (Illustrative short expansions of various intermediate quantities in 
the procedure can be found in Section III of \cite{KavaETC17}.)  Given eccentric orbital motion, the PN expansion 
necessarily involves an expansion in powers of eccentricity $e$ as well.  We have sought to go as deeply as 
possible in PN order and (especially) eccentricity order.  High-order expansion in eccentricity opens up the 
possibility of finding eccentricity-dependent terms that have closed-form expressions or infinite series with 
analytically-known coefficient sequences \cite{MunnEvan19a,MunnEvan20a,MunnEvan22a}.

We restate for emphasis one issue with calculating the PN and eccentricity expansions of the spin-precession 
invariant.  As we have discussed, the procedure begins with calculating the mode functions $X^\pm_{lmn}$ and their 
normalizations $C_{lmn}^\pm$.  Assume that $C_{lmn}^+$, for example, has been calculated in the general-$l$ case 
to a relative PN order of $\mathcal{P}$ and to order $e^{2 \mathcal{N}}$ in eccentricity.  We find that once 
derivatives have been taken, to calculate the metric perturbation and the self-force, and the various projections have 
been made, the expansion of the spin-precession invariant has lost two relative orders in $\mathcal{P}$ 
and two orders in $e^2$ (i.e., two order in $\mathcal{N}$).  For example, if $C_{lmn}^+$ were known in the 
general-$l$ case to 8PN (relative) and $e^{20}$, the computation of the general-$l$ contribution to the spin 
invariant is limited to 6PN relative order (7PN absolute order, as it is conventionally defined) and $e^{16}$.  
One caveat, however, is that the general-$l$ expansion does not contribute to the spin precession at half-integer 
orders.  Thus, as long as the specific-$l$ contributions are appropriately extended, the final result in this example 
would be an expansion for $\psi$ that reaches 7.5PN (absolute) and $e^{16}$.

\subsection{Regularization}

The final step in the procedure is regularization.  As mentioned in Sec.~\ref{sec:self-force}, instead of 
regularizing the self-force itself, we calculate $l$-modes of the spin invariant using $l$-modes of the 
unregularized retarded self-force.  Then, we make a mode-sum regularization of the spin invariant itself, in a 
procedure that is similar to but slightly more involved than the way in which we previously regularized the 
redshift invariant \cite{MunnEvan22a}.  Because the spin invariant involves derivatives of the metric perturbation, 
$l$-mode contributions to the singular behavior grow like $l$.  Mode-sum regularization requires subtracting off 
terms from the expansion of the singular field
\be
\label{eqn:modesum}
\D \psi = \sum_{l=0}^\infty \left( \D \psi^{l,\pm}_{\rm ret} \mp  A_S (2 l + 1) - B_S \right) .
\ee
The calculation on the right is direction dependent, based on whether the particle location is approached from 
outside or inside of $r_p$, but yields the same final value.  The coefficients, $A_S$ and $B_S$, are the two
regularization parameters required to allow the sum to converge.  However, usually regularization 
parameters are defined by decomposing every vector and tensor component into a sum over scalar spherical harmonics.  
Then, each regularization parameter is $l$ independent.  In the present application, our $l$ is different and refers 
to the tensor spherical harmonic index, with the implication \cite{WardWarb15} that $A_S$ at least is not completely 
independent of $l$ but changes value for $l=0$ and $l=1$.  Furthermore, neither $A_S$ nor $B_S$ are known a priori.

The workaround for the latter issue involves using the general-$l$ expansion.  Because $A_S$ and $B_S$ must match 
the large-$l$ behavior of $\D \psi^{l, \rm ret}$, we can expand our general-$l$ result for $\D \psi^{l}_{\rm ret}$ 
about $l=\infty$ to find two coefficients, $A_{\infty}$ and $B_{\infty}$.  Then theoretically, $B_{\infty} = B_S$ 
(for all $l$) and $A_{\infty} = A_S$ for $l \ge 2$.  The qualification is that regularization and decomposition 
of the singular field is usually discussed in the context of Lorenz gauge \cite{DetwWhit03}.  While we are working 
in a different gauge, mode-sum regularization carries across \cite{BaraOri01} to select other gauges, with RWZ 
gauge \cite{ThomWardWhit19} being one.  

The remaining problem is the behavior of $A_S$ when $l = 0,1$.  The solution is to recognize 
that $A_S$ flips sign across the location of the point mass \cite{WardWarb15} (see \eqref{eqn:MPchi}), allowing 
\eqref{eqn:modesum} to be replaced by \cite{KavaETC17,BiniDamoGera18} 
\be
\label{eqn:regComb}
\D \psi = \sum_{l=0}^\infty \left(\frac{1}{2} (\D\psi^{l, +}_{\rm ret} + \D\psi^{l, -}_{\rm ret}) - B_\infty \right) .
\ee
However, there is no reason to use \eqref{eqn:regComb} for every $l$.  We can instead use \eqref{eqn:modesum} 
for all $l \ge 2$ and reserve use of \eqref{eqn:regComb} for only the $l=0$ and $l=1$ modes.  The net effect is 
to reduce the calculation by roughly 50\%.

\begin{widetext}
\section{PN Expansions of the Spin-Precession Invariant}
\label{sec:PNresults}

The procedure in the previous two sections was utilized to compute the spin-precession invariant expansion to 9PN 
order and to $e^{16}$ in eccentricity.  We present the expansions in this section using both $1/p$ and $y$ as 
compactness parameters.  The expansions are given in this paper to 8PN order, with the full series being available 
in electronic form at the Black Hole Perturbation Toolkit \cite{BHPTK18} website and at our research group 
website \cite{UNCGrav22}.  

\subsection{Spin-precession invariant as an expansion in $1/p$}

Previous work in the circular-orbit limit has revealed \cite{AkcaDempDola17} the general PN structure of $\D\psi$ to 
9.5PN order.  Expressed in terms of the compactness parameter $1/p$, the form of the expansion is
\begin{align}
\label{eqn:psiInOneOverp}
\D \psi =& \,\, \frac{\D \psi_1^p}{p} + \frac{\D \psi_2^p}{p^2} +  \frac{\D \psi_3^p}{p^3} + \left(\D \psi_4^p 
+ \D \psi_{4L}^p \log p \right) \frac{1}{p^4} + \left(\D \psi_5^p + \D \psi_{5L}^p \log p \right) \frac{1}{p^5}  
+ \frac{\D \psi_{11/2}^p}{p^{11/2}}    \notag \\ &
+ \left(\D \psi_6^p + \D \psi_{6L}^p \log p \right) \frac{1}{p^6} + \frac{\D \psi_{13/2}^p}{p^{13/2}} 
+ (\D \psi_7^p + \D \psi_{7L}^p \log p  + \D \psi_{7L2}^p \log^2 p ) \frac{1}{p^7}  + \frac{\D \psi_{15/2}^p}{p^{15/2}} 
\notag \\ &
+ \left(\D \psi_8^p + \D \psi_{8L}^p \log p + \D \psi_{8L2}^p \log^2 p \right) \frac{1}{p^8}  
+ \left(\D \psi_{17/2}^p + \D \psi_{17/2L}^p \log p \right) \frac{1}{p^{17/2}}   \notag \\ &
+ \left(\D \psi_9^p + \D \psi_{9L}^p \log p + \D \psi_{9L2}^p \log^2 p \right) \frac{1}{p^9}  + \cdots .
\end{align}
Because this is a first-order self-force result, the entire right hand side should be viewed as multiplied by a 
factor of $\mu/M$.  For eccentric orbits, each of the quantities $\D\psi^p_k$, for different $k$, is no longer 
just a number but rather a function of the eccentricity $e$.  The purpose of this section is to show the form of 
these functions.  

In discussing energy and angular momentum fluxes (see e.g., \cite{MunnETC20}) it is conventional to factor out the 
circular-orbit limit with $p^{-5}$ and refer to terms in the expansion by their relative PN order.  Thus, the 
Peters-Mathews flux is 0PN relative.  In the spin-precession invariant expansion, the leading-order term is 
$p^{-1}$ and we will refer to this as the 1PN (absolute) term.  In our nomenclature, a $k$PN term is one that is
proportional to $p^{-k}$.

The first three functions were all found previously \cite{AkcaDempDola17} and shown to have closed forms
\begin{align}
\label{eqn:1PN}
\D \psi_1^p =& -1  ,   \\
\label{eqn:2PN}
\D \psi_2^p =& \frac{9 + 4 e^2}{4}  , \\
\label{eqn:3PN}
\D \psi_3^p =&  \left(\frac{739}{16}-\frac{123 \pi ^2}{64}\right)
+e^2 \left(\frac{341}{16}-\frac{123 \pi ^2}{256}\right)-\frac{e^4}{2}  .
\end{align}

Beyond 3PN order, only the circular-orbit behavior \cite{AkcaDempDola17} and first term, $e^2$, in eccentricity 
were known previously.  The $e^2$ terms through 6PN had been calculated by \cite{KavaETC17} and \cite{BiniDamoGera18} 
had found the $e^2$ behavior through 9PN.  Our present work extends every term through 9PN order to $e^{16}$ in 
eccentricity, allowing key parts of the functions $\D\psi^p_k(e)$ to be isolated in many cases and allowing us 
to find in a few cases complete, closed-form expressions.

The 4PN term is a case in which having enough terms in the eccentricity expansion allows us to identify elemental
parts of the eccentricity functions.  The behavior of the 4PN spin-precession invariant is reminiscent of the 
3PN energy flux.  We find first that the 4PN log term has a closed-form expression.  Then, that same function 
reappears in the 4PN non-log term.  It is then possible to see a grouping, as a series, that contains all of the 
log-transcendental numbers (which we denote by $\D\psi^{p,\chi}_4$ in analogy to similar functions in the energy flux, 
angular momentum flux, and redshift invariant expansions).  We display $\D\psi^{p,\chi}_4$ in this paper to $e^{16}$.  
The remaining part of the 4PN non-log term is a polynomial on the appearance of $\pi^2$ and a remaining 
rational-number series.  That latter rational-number series is also displayed here to $e^{16}$.  The breakdown of 
the 4PN term is as follows:
\begin{align}
\label{eqn:4PN}
\D \psi_4^p =& \bigg( -\frac{587831}{2880}-\frac{37961 e^2}{160}-\frac{28129 e^4}{480}-\frac{19015 e^6}{1152}
-\frac{138247 e^8}{15360}-\frac{12431 e^{10}}{2048}   \notag \\&
-\frac{327985 e^{12}}{73728} -\frac{56393 e^{14}}{16384}  -\frac{725137 e^{16}}{262144} + \cdots \bigg) 
+ \pi^2 \bigg( \frac{31697}{6144}-\frac{23729 e^2}{4096}-\frac{23761 e^4}{16384} \bigg)   \notag \\&
- 2 \left[\g_E +\log \left(\frac{8 (1-e^2)^{3/2}}{1+\sqrt{1-e^2}}\right) \right]  \D \psi_{4L}^p  
+ \D \psi_{4}^{p,\chi} , \\
\label{eqn:4PNchi}
\D \psi_{4}^{p,\chi}  =& \left(-\frac{2216 \log (2)}{15} +\frac{729 \log (3)}{5}\right)
+\left(\frac{55384 \log (2)}{15}-\frac{10206 \log (3)}{5}\right) e^2 +\bigg(-\frac{205917 \log (2)}{5}  +   \notag \\&
\frac{3620943 \log (3)}{320} + \frac{1953125 \log (5)}{192}\bigg) e^4
+\left(\frac{11518508 \log (2)}{45}+\frac{3995649 \log (3)}{320}-\frac{68359375 \log (5)}{576}\right) e^6  \notag \\&
+\left(-\frac{1597223897 \log (2)}{1350}-\frac{199689627159 \log (3)}{409600}
+\frac{274244140625 \log (5)}{442368}+\frac{678223072849 \log (7)}{3686400}\right) e^8  \notag \\&
+\left(\frac{27567693977 \log (2)}{4500}+\frac{1474172887599 \log (3)}{512000}
-\frac{71310546875 \log (5)}{36864}-\frac{678223072849 \log (7)}{307200}\right) e^{10}  \notag \\&
+\bigg(-\frac{39584616236117 \log (2)}{1134000}-\frac{2523359744732097 \log (3)}{458752000}
+\frac{1806540009765625 \log (5)}{445906944}  \notag \\&
+\frac{77643655602826369 \log (7)}{6370099200}\bigg) e^{12}
+\bigg(\frac{7796904819020377 \log (2)}{47628000}-\frac{96876093468033783 \log (3)}{3211264000}  \notag \\&
-\frac{173056322265625 \log (5)}{3121348608}-\frac{1313351173426101691 \log (7)}{31850496000}\bigg) e^{14}
+\bigg(-\frac{1349495913968063023 \log (2)}{2286144000}  \notag \\&
+\frac{44537688184006902231 \log (3)}{164416716800}  
-\frac{131695624109951171875 \log (5)}{1826434842624}  \notag \\&
+\frac{9397785802951526436547 \log (7)}{97844723712000}   
+\frac{81402749386839761113321 \log (11)}{4794391461888000}\bigg) e^{16} + \cdots , \\
\label{eqn:4PNL}
\D \psi_{4L}^p =& - \left( \frac{628}{15}+\frac{268 e^2}{5}+\frac{37 e^4}{10} \right)  .
\end{align}
Our result for the 4PN term is exceptional in one regard.  As mentioned in Sec.~\ref{sec:metPerExps}, we are able 
to exploit a feature in the eccentricity expansion of the general-$l$ part of the metric perturbation.  At any 
given PN order, the eccentricity expansion of the general-$l$ modes truncates at some power, which depends upon 
metric component and PN order but not $l$.  The only contributions to higher powers of $e$ beyond this truncation 
point come from the (MST-derived) specific-$l$ calculation.  We used this feature to calculate the 4PN term to much 
higher order in eccentricity ($e^{30}\, $!).  The resulting eccentricity series gave us another opportunity to look 
for additional closed-form expressions and infinite series with analytically recognizable coefficient sequences.  
While no additional such functions were identified, we are providing the full 4PN term to $e^{30}$ in the online 
repositories \cite{BHPTK18,UNCGrav22}.

The breakdown of the 5PN term is similar.  Once again, we have enough information in the lengthy eccentricity 
expansion to see that the 5PN log term is a polynomial.  The 5PN log term reappears in the 5PN non-log term.  
There is then a $\chi$-like grouping of terms that can be isolated in the 5PN non-log function.  There is a 
closed-form expression identifiable that multiplies $\pi^2$ and the remainder is a rational-number series.
\begin{align}
\label{eqn:5PN}
\D \psi_5^p =& \bigg( -\frac{48221551}{19200}-\frac{948244847 e^2}{403200}+\frac{213024509 e^4}{134400}
+\frac{6416801 e^6}{19200}+\frac{21598637 e^8}{161280}+\frac{61987887 e^{10}}{716800}  \notag \\&
+\frac{2519343 e^{12}}{40960}+\frac{482228861 e^{14}}{10321920}+\frac{25511929 e^{16}}{688128} 
+ \cdots \bigg) + \pi^2 \bigg( \frac{2483157}{8192}+\frac{21274445 e^2}{49152}     \\&
-\frac{1392915 e^4}{65536}  
-\frac{322801 e^6}{65536}-\frac{123}{256} \left(1-e^2\right)^{5/2} \bigg) 
- 2 \left[\g_E +\log \left(\frac{8 (1-e^2)^{3/2}}{1+\sqrt{1-e^2}}\right) \right]  \D \psi_{5L}^p  
+ \D \psi_{5}^{p,\chi}  ,  \notag \\
\label{eqn:5PNchi}
\D \psi_{5}^{p,\chi}   =& \bigg(\frac{155894 \log (2)}{105}-\frac{31347 \log (3)}{28}\bigg)
+\bigg(-\frac{4518706 \log (2)}{105} +\frac{4430133 \log (3)}{320}+\frac{9765625 \log (5)}{1344}\bigg) e^2 \notag \\&
+\bigg(\frac{103760293 \log (2)}{180}-\frac{31627665 \log (3)}{1792}  
-\frac{3850703125 \log (5)}{16128}\bigg) e^4+\bigg(-\frac{7257966409 \log (2)}{1512}  - \notag \\&
\frac{412866263889 \log (3)}{286720}  
+\frac{185751484375 \log (5)}{73728}+\frac{96889010407 \log (7)}{221184}\bigg) e^6
+\bigg(\frac{1363049747783 \log (2)}{37800}  \notag \\&
+\frac{193316307253281 \log (3)}{11468800} 
-\frac{179678368984375 \log (5)}{12386304}-\frac{458460436354739 \log (7)}{44236800}\bigg) e^8  \notag \\&
+\bigg(-\frac{12831673840577 \log (2)}{45360}-\frac{7224124751440749 \log (3)}{91750400}  
+\frac{8043417607578125 \log (5)}{148635648}   +   \notag \\&
\frac{355523097835854137 \log (7)}{3538944000}\bigg)e^{10}
+\bigg(\frac{183330830280580517 \log (2)}{95256000}
-\frac{805502573948046501 \log (3)}{12845056000}  \notag \\&
-\frac{422974129832265625 \log (5)}{4161798144}
-\frac{14424400165163864701 \log (7)}{25480396800}\bigg) e^{12}   \notag \\&
+\bigg(-\frac{4506759544919422111 \log (2)}{444528000}   
+\frac{4528120073662511265537 \log (3)}{1438646272000}   \notag \\&
-\frac{12939671783145268046875 \log (5)}{22373826822144}  
+\frac{17394718021348645585769 \log (7)}{8153726976000}  \notag \\&
+\frac{81402749386839761113321 \log (11)}{559345670553600}\bigg) e^{14}
+\bigg(\frac{8175474265144902875339 \log (2)}{192036096000}   \notag \\&
-\frac{487753344707796027586053 \log (3)}{23018340352000}
+\frac{8271291205691128966015625 \log (5)}{1073943687462912}    \\&
-\frac{6796075855660208932297721 \log (7)}{1174136684544000}
-\frac{1236467363350808533619347277 \log (11)}{402728882798592000}\bigg) e^{16} + \cdots ,  \notag \\
\label{eqn:5PNL}
\D \psi_{5L}^p =& \, \, \left(\frac{11153}{35}+\frac{11341 e^2}{15}+\frac{46467 e^4}{280}-\frac{1119 e^6}{560} \right).
\end{align}

Like the redshift invariant, the first half-integer function appears at 5.5PN order.  We find it to be a 
rational-number infinite series, multiplied by an overall factor of $\pi$
\begin{align}
\label{eqn:5p5PN}
\D \psi_{11/2}^p =& \, \, \pi \bigg( \frac{49969}{315}+\frac{319609 e^2}{630}+\frac{21280909 e^4}{100800}
+\frac{2619467 e^6}{362880}-\frac{5582939 e^8}{580608000}+\frac{19566341 e^{10}}{5806080000} \notag \\&
-\frac{1283076269 e^{12}}{2601123840000}-\frac{3498178499 e^{14}}{21849440256000}
+\frac{4868320009201 e^{16}}{251705551749120000} + \cdots \bigg) .
\end{align}

At 6PN order we found a structure similar to 4PN and 5PN, but with some added complexity.  Once again, we were
able to find a closed-form expression for the (6PN) log term, though it is not simply a polynomial.  The 6PN 
log term reappears in the 6PN non-log function.  We group the log-transcendental number terms into a $\chi$-like 
series again.  Then, an added wrinkle is the appearance of a $\pi^4$ term (which is a polynomial in $e$) as well 
as a more complicated closed-form expression multiplying $\pi^2$.  The remainder is, again, a rational-number 
series.  This breakdown is given by
\begin{align}
\label{eqn:6PN}
\D \psi_{6}^p =& \, \, \bigg( -\frac{1900873914203}{101606400}-\frac{465224579689 e^2}{5080320}
-\frac{2021344615177 e^4}{33868800}-\frac{481812394033 e^6}{203212800}  \notag \\&
+\frac{908657975293 e^8}{1625702400} +\frac{115787009753 e^{10}}{464486400} 
+\frac{63855468847 e^{12}}{371589120} + \frac{663864852377 e^{14}}{5202247680}   \notag \\&
+\frac{237445545371 e^{16}}{2378170368} + \cdots \bigg)  
+ \pi^4 \bigg( -\frac{7335303}{131072} -\frac{146026515 e^2}{1048576}-\frac{17998485 e^4}{524288}  
+\frac{679545 e^6}{16777216} \bigg)   \notag \\&
+ \pi^2 \bigg[ \frac{7254777827}{2359296} +\frac{32034966215 e^2}{2359296} +\frac{77315025809 e^4}{9437184}   
+\frac{5875228633 e^6}{12582912}  -  \frac{326041715 e^8}{33554432}  \notag \\&
+\left(-\frac{21405}{2048}-\frac{26549 e^2}{8192}\right) \left(1-e^2\right)^{5/2} \bigg]  
- 2 \left[\g_E +\log \left(\frac{8 (1-e^2)^{3/2}}{1+\sqrt{1-e^2}}\right) \right] \D \psi_{6L}^p 
+ \D \psi_{6}^{p,\chi} , \\
\label{eqn:6PNchi}
\D \psi_{6}^{p,\chi} =& \, \, \bigg[\bigg(-\frac{5637649 \log (2)}{630} +\frac{234009 \log (3)}{70}
+\frac{9765625 \log (5)}{9072}\bigg)  +\bigg(\frac{278347639 \log (2)}{945}
-\frac{159335343 \log (3)}{8960}   \notag \\&
-\frac{17193359375 \log (5)}{145152}\bigg) e^2   
+\bigg(-\frac{67298137969 \log (2)}{15120}-\frac{6160676211 \log (3)}{4480}
+\frac{2073211184375 \log (5)}{870912}   \notag \\&
+\frac{96889010407 \log (7)}{248832}\bigg) e^4    
+\bigg(\frac{45349640544529 \log (2)}{816480}+\frac{31165257813381 \log (3)}{1146880}    \notag \\&
-\frac{182206079940625 \log (5)}{7962624}-\frac{387081765684929 \log (7)}{23887872}\bigg) e^6  
+\bigg(-\frac{558659555193209 \log (2)}{816480}    \notag \\&
-\frac{53870960843541 \log (3)}{262144}
+\frac{61171359261321875 \log (5)}{445906944}   
+\frac{294158696797354949 \log  (7)}{1194393600}\bigg) e^8   \notag \\&
+\bigg(\frac{847397764060731841 \log (2)}{122472000}+\frac{32809717319486013 \log (3)}{1835008000} 
-\frac{2422221742760078125 \log (5)}{5350883328}  \notag \\&
-\frac{2408276754022760847373 \log (7)}{1146617856000}\bigg) e^{10}
+\bigg(-\frac{16872974428064804681 \log (2)}{321489000}    \notag \\&
+\frac{702548354497142812293 \log (3)}{51380224000}  
-\frac{10596237215626599434375 \log (5)}{5393690394624}    \notag \\&
+\frac{16313635952096120303239 \log (7)}{1375941427200}  
+\frac{81402749386839761113321 \log (11)}{134842259865600}\bigg) e^{12}   \notag \\&
+\bigg(\frac{44527790291197983392407 \log (2)}{144027072000}  
-\frac{796933093810646955107709 \log (3)}{5754585088000}    \notag \\&
+\frac{110679665407920978635528125 \log (5)}{2416373296791552}  
-\frac{42403949775722831599753919 \log (7)}{880602513408000}    \notag \\&
-\frac{151763386707026658481899113 \log (11)}{8629904631398400}\bigg) e^{14}  
+\bigg(-\frac{1066795518870736337912533 \log (2)}{691329945600}    \notag \\&
+\frac{561883567881441059477453079 \log (3)}{736586891264000}  
-\frac{68528286383578337812136246875 \log (5)}{173978877368991744}   \notag \\&
+\frac{18715907817618872220720781253 \log (7)}{126806761930752000}   
+\frac{624926196399309721148911875383 \log (11)}{2718419958890496000}  \notag \\&
+\frac{91733330193268616658399616009 \log (13)}{8698943868449587200}\bigg) e^{16} + \cdots  \bigg]
, \\
\label{eqn:6PNL}
\D \psi_{6L}^p =& \, \, \frac{454397}{3780}+\frac{2384929 e^2}{1890}+\frac{45143023 e^4}{30240}
+\frac{5037481 e^6}{20160}-\frac{2387 e^8}{960} - \left(\frac{146}{5} + \frac{37 e^2}{5}\right) 
\left(1-e^2\right)^{5/2} .
\end{align}

The 6.5PN term is similar in form to the 5.5PN term, a rational-number infinite series with an overall factor of 
$\pi$
\begin{align}
\label{eqn:6p5PN}
\D \psi_{13/2}^p =& \, \, \pi \bigg(-\frac{2620819}{2100}-\frac{1586616631 e^2}{235200}
-\frac{5337465431 e^4}{940800} -\frac{273779628487 e^6}{406425600}    \notag \\&
+\frac{23921441479 e^8}{5419008000}  
-\frac{2200323829 e^{10}}{19267584000}  
+\frac{6113757813097 e^{12}}{655483207680000}   \notag \\&
+\frac{200661326003101 e^{14}}{48942746173440000}
-\frac{13890607636693243 e^{16}}{7047755448975360000}  + \cdots \bigg)  .
\end{align}

At 7PN order, there are additional complexities.  Here, a $\log^2p$ term makes its first appearance and we find the 
closed-form expression for that term.  The 7PN log term then inherits the general structure of the 4PN non-log 
term.  It features a reappearance of the 7PN $\log^2p$ term, a $\chi$-like series grouping, and a remaining 
rational-number series.  The 7PN non-log term is the first appearance of a much more complicated expansion, 
which features numerous transcendental number terms.  While we have calculated it to $e^{16}$, it is sufficiently 
complicated that we only present the first few coefficients here.  The entire term is available online 
\cite{BHPTK18,UNCGrav22}.  The breakdown of the 7PN terms is
\begin{align}
\label{eqn:7PN}
\D \psi_{7}^p =& \, \, \bigg( -\frac{1282190594044678657}{7041323520000}
+\frac{1316474014843 \g_E}{43659000}-\frac{3396608 \g_E^2}{1575}  
+\frac{25657561505749 \pi ^2}{2477260800}    \notag \\&
+\frac{341587582057 \pi ^4}{1006632960}  
+\frac{2783260080883 \log (2)}{43659000}-\frac{5149696 \g_E \log (2)}{1575}  
-\frac{931328 \log ^2(2)}{1575}    \notag \\&
+\frac{282123979047 \log (3)}{8624000}   
-\frac{936036}{175} \g_E \log (3) - \frac{936036}{175} \log (2) \log (3)-\frac{468018 \log ^2(3)}{175}
+\frac{63488 \zeta (3)}{15}  \notag \\&
-\frac{361328125 \log (5)}{24192}   \bigg) + \bigg( -\frac{2888955324477314921}{2347107840000}  
+\frac{1075057978433 \g_E}{4851000} - \frac{7219504 \g_E^2}{525}   \notag \\&
+\frac{465082867177871 \pi ^2}{6606028800} +\frac{314165501411 \pi ^4}{335544320}
+\frac{2387982140729 \log (2)}{8731800} - \frac{79652512}{315} \g_E \log (2)   \notag \\&
-\frac{80263696 \log ^2(2)}{175} -\frac{299782486660473 \log (3)}{275968000}
+\frac{15912612}{175} \g_E \log (3) +\frac{15912612}{175} \log (2) \log (3)    \notag \\&
+\frac{7956306 \log ^2(3)}{175} 
+\frac{2411543359375 \log (5)}{2838528}+\frac{678223072849 \log (7)}{6082560}
+\frac{134944 \zeta (3)}{5}  \bigg) e^2      \notag \\&
+ \bigg(  -\frac{113070994466917771}{58677696000} 
+\frac{1472132397523 \g_E}{4656960} -\frac{1627684 \g_E^2}{105}
+\frac{1927756649323 \pi ^2}{18350080}      \notag \\&
-\frac{230109551791 \pi ^4}{268435456}  
+\frac{4388818385774297 \log (2)}{349272000} 
+\frac{1697626904}{525} \g_E\log (2)+\frac{3404199436 \log ^2(2)}{525}    \notag \\&
+\frac{157855832267967 \log (3)}{7168000}  
-\frac{262324089}{400} \g_E \log (3)   
-\frac{262324089}{400} \log (2) \log (3)-\frac{262324089 \log ^2(3)}{800}    \notag \\&
-\frac{7905614284523125 \log (5)}{1072963584}-\frac{1044921875 \g_E \log (5)}{1008}  
-\frac{1044921875 \log (2) \log (5)}{1008}   \notag \\&
-\frac{1044921875 \log ^2(5)}{2016}
-\frac{1107664826873969 \log (7)}{109486080}+30424 \zeta (3) \bigg) e^4  + \cdots 
, \\
\label{eqn:7PNL}
\D \psi_{7L}^p =& \, \, \bigg( -\frac{1316474014843}{87318000}-\frac{1048904096833 e^2}{9702000}
-\frac{6494255602463 e^4}{46569600}-\frac{2123271392639 e^6}{55883520}  \notag \\&
-\frac{1333068673 e^8}{352800} - \frac{148643207731 e^{10}}{88704000}
-\frac{5888596871 e^{12}}{5376000}-\frac{204908923 e^{14}}{258048}    \notag \\&
-\frac{63078518299 e^{16}}{103219200}  + \cdots \bigg)
- 4 \left[\g_E +\log \left(\frac{8 (1-e^2)^{3/2}}{1+\sqrt{1-e^2}}\right) \right] 
\D \psi_{7L2}^p + \D \psi_{7L}^{p,\chi}  , \\
\label{eqn:7PNLchi}
\D \psi_{7L}^{p,\chi} =& \, \, \bigg(-\frac{602624 \log (2)}{225} + \frac{468018 \log (3)}{175}\bigg)
+\bigg(\frac{155814256 \log (2)}{1575} -\frac{7956306 \log (3)}{175}\bigg) e^2  \notag \\&
+\bigg(-\frac{865090292 \log (2)}{525} +\frac{262324089 \log (3)}{800}+\frac{1044921875 \log(5)}{2016}\bigg) e^4 
+\bigg(\frac{7561620022 \log (2)}{525}    \notag \\&
+\frac{418174083 \log (3)}{200} -\frac{11494140625 \log(5)}{1512}\bigg) e^6 
+ \bigg(-\frac{27112112493049 \log (2)}{283500}    \notag \\&
- \frac{339134405423319 \log (3)}{7168000}  + \frac{238205615234375 \log (5)}{4644864}   
+\frac{507989081563901 \log (7)}{27648000}\bigg) e^8      \notag \\&
+ \bigg(\frac{2438441188502 \log (2)}{3375}  
+\frac{2457755729168913 \log (3)}{7168000}-\frac{328215185546875 \log (5)}{1548288}  -  \notag \\&
 \frac{507989081563901 \log (7)}{1843200}\bigg) e^{10}   
+\bigg(-\frac{27531157668664681 \log (2)}{4961250}
-\frac{6481474893571248729 \log (3)}{8028160000}   \notag \\&
+\frac{2822922907568359375 \log (5)}{4682022912}
+\frac{92389498231271366573 \log (7)}{47775744000}\bigg) e^{12}  +  \notag \\&
 \bigg(\frac{20936989659899360021 \log (2)}{625117500}  
-\frac{171956257323847250841 \log (3)}{28098560000}
-\frac{510786718173828125 \log (5)}{16387080192}   \notag \\&
-\frac{1009210224594126146977 \log (7)}{119439360000}\bigg) e^{14}  
+\bigg(-\frac{4619194353324708185237 \log (2)}{30005640000}  \notag \\&
+\frac{982128291923792960826417 \log (3)}{14386462720000}   
-\frac{2383328610155127138671875 \log (5)}{134242960932864}    \\&
+\frac{18946304992805061334986887 \log (7)}{733835427840000}   
+\frac{1053921396311414387134166987 \log (11)}{251705551749120000}\bigg) e^{16} + \cdots  \notag
, \\
\label{eqn:7PNL2}
\D \psi_{7L2}^p =& \, \, -\frac{849152}{1575}-\frac{1804876 e^2}{525}-\frac{406921 e^4}{105}
-\frac{543667 e^6}{630}-\frac{10593 e^8}{560}  .
\end{align}

The 7.5PN term is a rational-number infinite series (multiplied by an overall factor of $\pi$), like the two 
half-integer contributions before it
\begin{align}
\label{eqn:7p5PN}
\D \psi_{15/2}^p =& \, \, \pi \bigg( \frac{2782895449}{2910600}+\frac{2404331748779 e^2}{279417600}
+\frac{76598649855971 e^4}{6035420160}+\frac{8047650291899 e^6}{5267275776}    \notag \\&
-\frac{5292862534207931 e^8}{10534551552000}-\frac{279109475326162289 e^{10}}{27811216097280000}
-\frac{4713058120092455293 e^{12}}{2336142152171520000}  \notag \\&
-\frac{959696950970026746929 e^{14}}{523295842086420480000}
-\frac{25409776245923969873 e^{16}}{37378274434744320000} + \cdots \bigg)  .
\end{align}

The description of the breakdown in the 8PN eccentricity functions is essentially the same as what we said 
about the 7PN terms just prior to \eqref{eqn:7PN}.  We still find a polynomial for the 8PN $\log^2p$ term, albeit 
one order in $e^2$ longer.  Because of the complexity of the 8PN non-log term, we give it only though $e^4$ and 
leave the full expansion though $e^{16}$ to the online repositories \cite{BHPTK18,UNCGrav22}.  The 8PN 
spin-precession-invariant correction splits into
\begin{align}
\label{eqn:8PN}
\D \psi_{8}^p 
=& \bigg(\frac{78550205239878250993769}{28193459374080000}-\frac{1888832198890393 \g_E }{15891876000}
+\frac{177306208 \g_E^2}{11025}+\frac{569460279231731 \pi ^2}{123312537600}  \notag \\&
-\frac{623848083842333 \pi ^4}{21474836480}-\frac{41942063811247 \log (2)}{1059458400}
+\frac{2720192 \g_E  \log (2)}{11025}-\frac{520925728 \log ^2(2)}{11025}  \notag \\&
-\frac{868469344973829 \log (3)}{3139136000}+\frac{59742279 \g_E  \log (3)}{1225}
+\frac{59742279 \log (2) \log (3)}{1225}+\frac{59742279 \log^2(3)}{2450}  \notag \\&
+\frac{8570767578125 \log (5)}{96864768}
+\frac{678223072849 \log (7)}{92664000}-\frac{861696 \zeta (3)}{35}\bigg)  
+\bigg(\frac{12590844685671737819611}{939781979136000}    \notag \\&
-\frac{3978068608616891 \g_E }{2648646000}
+\frac{371280152 \g_E^2}{2205}-\frac{33310259864964463 \pi ^2}{443925135360}   
-\frac{31118085613898053 \pi^4}{257698037760}    \notag \\&
-\frac{88630614687481099 \log (2)}{7945938000}
+\frac{38342498672 \g_E  \log (2)}{11025}+\frac{7893208952 \log^2(2)}{1225}  
+\frac{812331139710343959 \log (3)}{100452352000}    \notag \\&
-\frac{621149553}{784} \g_E  \log (3) -\frac{621149553}{784} \log(2) \log (3)-\frac{621149553 \log ^2(3)}{1568}   
-\frac{196313675703125 \log (5)}{21697708032}    \notag \\&
-\frac{3173828125 \g_E  \log(5)}{7056} -\frac{3173828125 \log (2) \log (5)}{7056}
-\frac{3173828125 \log ^2(5)}{14112}  -\frac{55100101995388051 \log (7)}{23721984000}   \notag \\&
-\frac{5551264 \zeta (3)}{21}\bigg) e^2  
+\bigg(\frac{3657307066227250447201}{293681868480000}-\frac{159538901113146521 \g_E }{42378336000} 
+\frac{397837653 \g_E^2}{1225}   \notag \\&
-\frac{384067364086852189 \pi ^2}{2959500902400}  
-\frac{4915898447923097 \pi ^4}{85899345920}+\frac{82885306824453137 \log (2)}{54486432000}    \notag \\&
-\frac{246163497086 \g_E  \log (2)}{4725}  -\frac{482425731551 \log ^2(2)}{4725}   
-\frac{16338364486486698339 \log (3)}{200904704000}+\frac{15151367601 \g_E  \log (3)}{78400}   \notag \\&
-\frac{456484805199 \log (2) \log (3)}{78400}+\frac{15151367601 \log ^2(3)}{156800}    
-\frac{12490894332540370625 \log (5)}{130186248192}  \notag \\&
+\frac{2022790234375 \g_E  \log (5)}{84672}  
+\frac{2022790234375 \log (2) \log (5)}{84672}  +\frac{2022790234375 \log ^2(5)}{169344}    \notag \\&
+\frac{16777483050927098843 \log (7)}{142331904000}-\frac{17811754 \zeta (3)}{35}\bigg) e^4  + \cdots  
, \\
\label{eqn:8PNL}
\D \psi_{8L}^p =& \bigg(\frac{1884153630595993}{31783752000}+\frac{3866001511074491 e^2}{5297292000}
 +\frac{140691202491074201 e^4}{84756672000}+\frac{8446543158422249 e^6}{9246182400}   \notag \\&
 +\frac{2393628746500801 e^8}{16951334400}+\frac{8815849170404069 e^{10}}{226017792000}
 +\frac{37201967185037 e^{12}}{1549836288}+\frac{5703127370959 e^{14}}{338688000}   \notag \\&
 +\frac{6115396987321 e^{16}}{481689600}+ \cdots \bigg) 
 - 4 \left[\g_E +\log \left(\frac{8 (1-e^2)^{3/2}}{1+\sqrt{1-e^2}}\right) \right] 
\D \psi_{8L2}^p + \D \psi_{8L}^{p,\chi}  
, \\
\label{eqn:8PNLchi}
\D \psi_{8L}^{p,\chi} =&  \bigg(\frac{70650464 \log (2)}{2205}-\frac{59742279 \log (3)}{2450}\bigg)
+\bigg(-\frac{2208349688 \log (2)}{1575}+\frac{621149553 \log (3)}{1568}  \notag \\&
+\frac{3173828125 \log (5)}{14112}\bigg) e^2 
+\bigg(\frac{883055473063 \log (2)}{33075}-\frac{15151367601 \log (3)}{156800}
-\frac{2022790234375 \log (5)}{169344}\bigg) e^4   \notag \\&
+\bigg(-\frac{9735335606114 \log (2)}{33075}-\frac{2983923884637 \log (3)}{28672}
+\frac{2682838643359375 \log (5)}{16257024}    \notag \\&
+\frac{8816899947037 \log (7)}{331776}\bigg) e^6  
+\bigg(\frac{23310192268124059 \log (2)}{7938000}+\frac{302771958934455009 \log (3)}{200704000}   \notag \\&
-\frac{156689829056640625 \log (5)}{130056192}-\frac{299112464634116791 \log (7)}{331776000}\bigg) e^8
+ \cdots
, \\
\label{eqn:8PNL2}
\D \psi_{8L2}^{p} =& \frac{44326552}{11025}+\frac{92820038 e^2}{2205}+\frac{397837653 e^4}{4900}
+\frac{155002853 e^6}{4410}+\frac{332684587 e^8}{141120}-\frac{100899 e^{10}}{6272} .
\end{align}

\subsection{Spin-precession invariant as an expansion in $y$}

The compactness parameter $1/p$ is easily related to $e$ and the alternative compactness parameter $y$, written 
as a PN expansion.  That relationship allows \eqref{eqn:psiInOneOverp} to be recast as an expansion in 
$y$
\begin{align}
\D \psi =& \,\, \D \psi_1^y y + \D \psi_2^y y^2 +  \D \psi_3^y y^3 
+ \left(\D \psi_4^y + \D \psi_{4L}^y \log y \right) y^4  + \left(\D \psi_5^y + \D \psi_{5L}^y \log y \right) y^5  
+ \D \psi_{11/2}^y y^{11/2}    \notag \\ &
+ \left(\D \psi_6^y + \D \psi_{6L}^y \log y \right) y^6 
+ \D \psi_{13/2}^y y^{13/2} + \left(\D \psi_7^y + \D \psi_{7L}^y \log y + \D \psi_{7L2}^y \log^2 y \right) y^7
+ \D \psi_{15/2}^y y^{15/2}   \notag \\ &
+ \left(\D \psi_8^y + \D \psi_{8L}^y \log y + \D \psi_{8L2}^y \log^2 y \right) y^8
+ \left(\D \psi_{17/2}^y + \D \psi_{17/2}^y \log y \right) y^{17/2}   \notag \\ &
+ \left(\D \psi_9^y + \D \psi_{9L}^y \log y + \D \psi_{9L2}^y \log^2 y \right) y^9 + \cdots .
\end{align}
Again, the entire right hand side should be viewed as multiplied by a factor of $\mu/M$.  Because of the 
reparameterization, many of the functions $\D\psi^y_k(e)$ differ from their analogs $\D\psi^p_k(e)$ in the 
prior subsection.

The first three terms differ from \eqref{eqn:1PN}-\eqref{eqn:3PN} but still have simple closed forms
\begin{align}
\D \psi_1^y =& \, \, \frac{-1}{1-e^2}  ,   \\
\D \psi_2^y =& \,\, \frac{1}{(1-e^2)^2} \left(\frac{9}{4} + 3 e^2  \right)  , \\
\D \psi_3^y =& \,\, \frac{1}{(1-e^2)^3} \left(\frac{819}{16}-\frac{123 \pi ^2}{64}+e^2 \left(\frac{173}{16}
-\frac{123 \pi ^2}{256}\right)-\frac{15 e^4}{2} \right) - \frac{5}{(1-e^2)^{3/2}} .
\end{align}
Note the appearance of eccentricity singular factors of increasing power.

The rest of the terms closely mirror their counterparts in the $1/p$ expansion.  We will refer the reader back to 
the previous subsection for discussion.  The 4PN terms are similar in form to those in \eqref{eqn:4PN}, 
\eqref{eqn:4PNchi}, and \eqref{eqn:4PNL}
\begin{align}
\D \psi_4^y =& \,\,  \frac{1}{(1-e^2)^4} \bigg( -\frac{587831}{2880}-\frac{82781 e^2}{160}-\frac{99259 e^4}{480}
+\frac{15821 e^6}{1152}-\frac{100147 e^8}{15360}-\frac{10451  e^{10}}{2048}  \notag \\&
-\frac{291445 e^{12}}{73728} -\frac{51573 e^{14}}{16384}-\frac{674917 e^{16}}{262144} + \cdots  \bigg)
+ \frac{\pi^2}{(1-e^2)^4} \bigg(\frac{31697}{6144}+\frac{23503 e^2}{4096}+\frac{23471 e^4}{16384} \bigg) \notag \\&
+ 2 \left[\g_E +\log \left(\frac{8 (1-e^2)}{1+\sqrt{1-e^2}}\right) \right] \D \psi_{4L}^y + \D \psi_{4}^{y,\chi}  , \\
\D \psi_{4}^{y,\chi} =& \,\, \frac{1}{(1-e^2)^4} \bigg[ \left(\frac{729 \log (3)}{5} - \frac{2216 \log (2)}{15} \right)
+\left(\frac{55384 \log (2)}{15}-\frac{10206 \log (3)}{5}\right) e^2 +\bigg(\frac{3620943 \log (3)}{320}  -   \notag \\&
\frac{205917 \log (2)}{5}  + \frac{1953125 \log (5)}{192}\bigg) e^4
+\left(\frac{11518508 \log (2)}{45}+\frac{3995649 \log (3)}{320}-\frac{68359375 \log (5)}{576}\right) e^6  \notag \\&
+\left(\frac{274244140625 \log (5)}{442368}-\frac{1597223897 \log (2)}{1350}
-\frac{199689627159 \log (3)}{409600} +\frac{678223072849 \log (7)}{3686400}\right) e^8  \notag \\&
+\left(\frac{27567693977 \log (2)}{4500}+\frac{1474172887599 \log (3)}{512000}
-\frac{71310546875 \log (5)}{36864}-\frac{678223072849 \log (7)}{307200}\right)   \notag \\&  \times e^{10}
+\bigg(-\frac{39584616236117 \log (2)}{1134000}-\frac{2523359744732097 \log (3)}{458752000}
+\frac{1806540009765625 \log (5)}{445906944}  \notag \\&
+\frac{77643655602826369 \log (7)}{6370099200}\bigg) e^{12}
+\bigg(\frac{7796904819020377 \log (2)}{47628000}-\frac{96876093468033783 \log (3)}{3211264000} - \notag \\&
\frac{173056322265625 \log (5)}{3121348608}-\frac{1313351173426101691 \log (7)}{31850496000}\bigg) e^{14}
+\bigg(-\frac{1349495913968063023 \log (2)}{2286144000}  \notag \\&
+\frac{44537688184006902231 \log (3)}{164416716800}  
-\frac{131695624109951171875 \log (5)}{1826434842624}  \notag \\&
+\frac{9397785802951526436547 \log (7)}{97844723712000}   
+\frac{81402749386839761113321 \log (11)}{4794391461888000}\bigg) e^{16} + \cdots \bigg]   , \\
\D \psi_{4L}^y =& \,\, \frac{1}{(1-e^2)^4} \left( \frac{628}{15}+\frac{268 e^2}{5}+\frac{37 e^4}{10} \right) . 
\end{align}
The polynomial part of the 4PN log term is identical to that in \eqref{eqn:4PNL}, reflecting its nature as a 
leading-logarithm term.  Like the 4PN term in the $1/p$ expansion, we also calculated this term in the $y$ 
expansion to $e^{30}$, and that entire dependence is reproduced in the online repositories \cite{BHPTK18,UNCGrav22}.

The 5PN terms mirror those in \eqref{eqn:5PN}, \eqref{eqn:5PNchi}, and \eqref{eqn:5PNL}
\begin{align}
\D \psi_5^y =& \,\, \frac{1}{(1-e^2)^5} \bigg(-\frac{48221551}{19200}-\frac{218002469 e^2}{134400}
+\frac{554849699 e^4}{134400}+\frac{15403763 e^6}{9600}+\frac{45454391 e^8}{215040}  \notag \\&
+\frac{175906613 e^{10}}{1075200}+\frac{9382421 e^{12}}{81920}+\frac{147111571 e^{14}}{1720320}
+\frac{736852183  e^{16}}{11010048} + \cdots \bigg)  \notag \\&
+ \frac{\pi^2}{(1-e^2)^5} \bigg[ \frac{2719317}{8192}+\frac{6391663 e^2}{16384}-\frac{1142291 e^4}{65536}
-\frac{223697 e^6}{65536} - (1-e^2)^{3/2} \left(\frac{7503}{256} + \frac{861 e^2}{128} \right) \bigg]  \notag \\&
+ 2 \left[\g_E +\log \left(\frac{8 (1-e^2)}{1+\sqrt{1-e^2}}\right) \right] \D \psi_{5L}^y + \D \psi_{5}^{y,\chi} 
, \\
\D \psi_{5}^{y,\chi} 
=& \,\, \frac{1}{(1-e^2)^5} \bigg[ \bigg(\frac{155894 \log (2)}{105}-\frac{31347 \log (3)}{28}\bigg)
+\bigg(-\frac{292974 \log (2)}{7} +\frac{811377 \log (3)}{64}  \notag \\&
+\frac{9765625 \log (5)}{1344}\bigg) e^2   +\bigg(\frac{98443429 \log (2)}{180}-\frac{11825109 \log (3)}{8960}
-\frac{3850703125 \log (5)}{16128}\bigg) e^4 + \notag \\&
\bigg(\frac{179751484375 \log (5)}{73728}  -\frac{33799060013 \log (2)}{7560}
-\frac{438821183313 \log (3)}{286720}  +\frac{96889010407 \log (7)}{221184}\bigg) e^6    \notag \\&
+\bigg(\frac{1285645374023 \log (2)}{37800}+\frac{192170674772001 \log (3)}{11468800}   
-\frac{167918368984375 \log (5)}{12386304}  \notag \\&
-\frac{458460436354739 \log  (7)}{44236800}\bigg) e^8  
+\bigg(-\frac{62011700285317 \log (2)}{226800}-\frac{6866280939571821 \log (3)}{91750400}  +  \notag \\&
\frac{7306249357578125 \log (5)}{148635648}+\frac{350314344636373817 \log (7)}{3538944000}\bigg)e^{10}
+\bigg(\frac{178662406711739429 \log (2)}{95256000} - \notag \\&
\frac{1101374969180716197 \log (3)}{12845056000} 
-\frac{358568725832265625 \log (5)}{4161798144}
-\frac{13974363888728765053 \log (7)}{25480396800}\bigg) e^{12}  \notag \\&
+\bigg(-\frac{4382622188402959199 \log (2)}{444528000}   
+\frac{4591426122938350115073 \log (3)}{1438646272000}  \notag \\&
-\frac{4554943798461756015625 \log (5)}{7457942274048}    
+\frac{16599646987975703567209 \log (7)}{8153726976000}  \notag \\&
+\frac{81402749386839761113321 \log (11)}{559345670553600}\bigg) e^{14}  
+\bigg(\frac{7923977303302581594827 \log (2)}{192036096000}   \notag \\&
-\frac{482198082003965098333701 \log (3)}{23018340352000}  
+\frac{8271767545294840966015625 \log (5)}{1073943687462912}   \\&
-\frac{6408752834402770430401529 \log (7)}{1174136684544000}  
-\frac{1236467363350808533619347277 \log (11)}{402728882798592000}\bigg) e^{16} + \cdots \bigg]  , \notag \\
\D \psi_{5L}^y =& \,\, - \frac{1}{(1-e^2)^5} \bigg( \frac{11153}{35}+1091 e^2+\frac{166531 e^4}{280}
+\frac{15457 e^6}{560}  \bigg) .
\end{align}
The polynomial part of the 5PN log term differs from that in \eqref{eqn:5PNL}, as expected since it is a 
1PN-log (i.e., a 1PN correction to a leading log) \cite{MunnEvan20a}.

The 5.5PN (first half-integer PN) term is similar to its $1/p$ expansion \eqref{eqn:5p5PN}
\begin{align}
\D \psi_{11/2}^y =& \,\,  \frac{\pi}{(1-e^2)^{11/2}} \bigg( \frac{49969}{315}+\frac{319609 e^2}{630}
+\frac{21280909 e^4}{100800} + \frac{2619467 e^6}{362880}-\frac{5582939 e^8}{580608000}  \notag \\&
+\frac{19566341 e^{10}}{5806080000}-\frac{1283076269 e^{12}}{2601123840000}
-\frac{3498178499  e^{14}}{21849440256000}+\frac{4868320009201 e^{16}}{251705551749120000} 
+ \cdots \bigg) .
\end{align}

The 6PN term splits into parts that mimic \eqref{eqn:6PN}, \eqref{eqn:6PNchi}, and \eqref{eqn:6PNL} in the $1/p$ 
expansion of the 6PN term, with the exception of the appearance of eccentricity singular factors
\begin{align}
\D \psi_6^y =& \,\, \frac{1}{(1-e^2)^6}\bigg(-\frac{1900873914203}{101606400}-\frac{327006360319 e^2}{5080320}
-\frac{1043250935257 e^4}{33868800}-\frac{5158311322393 e^6}{203212800}   \notag \\&
-\frac{14506405082507 e^8}{1625702400}-\frac{869231282527 e^{10}}{464486400}  
-\frac{86093410741 e^{12}}{74317824}-\frac{4056226220383 e^{14}}{5202247680}   
-\frac{1350712174045 e^{16}}{2378170368}   \bigg) \notag \\&
+ \frac{\pi^4}{(1-e^2)^6} \bigg(-\frac{7335303}{131072}-\frac{146026515 e^2}{1048576}
-\frac{17998485 e^4}{524288}+\frac{679545 e^6}{16777216} \bigg) 
+ \frac{\pi^2}{(1-e^2)^6}  \bigg[ \bigg(\frac{215485}{6144}  \notag \\&
+\frac{1156631 e^2}{8192} +\frac{280939 e^4}{8192}\bigg)\left(1-e^2\right)^{3/2}
+\frac{7147373027}{2359296}+\frac{24774167687 e^2}{2359296} +\frac{38112094481 e^4}{9437184} \notag \\&
+\frac{7266864217 e^6}{12582912}  +\frac{445428621 e^8}{33554432} \bigg]  
+ 2 \left[\g_E +\log \left(\frac{8 (1-e^2)}{1+\sqrt{1-e^2}}\right) \right] \D \psi_{6L}^y 
+ \D \psi_{6}^{y,\chi}  \notag \\
\D \psi_{6}^{y,\chi}=& \frac{1}{(1-e^2)^6}\bigg[\bigg(-\frac{5637649 \log(2)}{630}
+\frac{234009 \log (3)}{70}+\frac{9765625 \log (5)}{9072}\bigg) +\bigg(\frac{38238253 \log (2)}{135}  \notag \\&
-\frac{18075555 \log (3)}{1792}-\frac{17193359375 \log (5)}{145152}\bigg) e^2 
+\bigg(-\frac{12445643645 \log (2)}{3024}  -\frac{6533011503 \log (3)}{4480}      \notag \\&
+\frac{2009929934375 \log (5)}{870912}  + \frac{96889010407 \log (7)}{248832}\bigg) e^4  
+ \bigg( \frac{8316214161581 \log (2)}{163296} + \frac{30954590856837 \log (3)}{1146880}    \notag \\&
- \frac{1155970259584375 \log (5)}{55738368}  
- \frac{387081765684929 \log (7)}{23887872}\bigg) e^6+\bigg(-\frac{525941334093929 \log (2)}{816480}   \notag \\&
-\frac{1751622570112671 \log (3)}{9175040} +\frac{51404505486321875 \log (5)}{445906944}  
+\frac{288926690235376949 \log (7)}{1194393600}\bigg) e^8   \notag \\&
+\bigg( \frac{808070130394854601 \log (2)}{122472000}  
-\frac{254004783448418307 \log (3)}{1835008000}-\frac{1753181559747578125 \log (5)}{5350883328}  \notag \\&
-\frac{2294506717029507369613 \log (7)}{1146617856000}\bigg) e^{10}  
+\bigg( -\frac{16027246190731123427 \log (2)}{321489000}  \notag \\&
+\frac{738364400868241156581 \log (3)}{51380224000}
-\frac{13119943727175549434375 \log (5)}{5393690394624}   \notag \\&
+\frac{15015279849631826906071 \log (7)}{1375941427200} 
+\frac{81402749386839761113321 \log (11)}{134842259865600}\bigg) e^{12}   \notag \\&
+\bigg( \frac{41914674050681714318023 \log (2)}{144027072000}  
-\frac{791847483991723599711741 \log (3)}{5754585088000}   \notag \\&
+\frac{112698947723278715435528125 \log (5)}{2416373296791552}  
-\frac{37760848394198321304112319 \log (7)}{880602513408000}   \notag \\&
-\frac{151763386707026658481899113 \log (11)}{8629904631398400}\bigg) e^{14}  
+\bigg(  
-\frac{5002336841855648660824877 \log (2)}{3456649728000}     \notag \\&
+\frac{539059140139095745481254167 \log (3)}{736586891264000} 
-\frac{67491579530903875816811246875 \log (5)}{173978877368991744}   \notag \\&
+\frac{16203069304365251842129022021 \log (7)}{126806761930752000}  
+\frac{620970022779109308758804474783 \log (11)}{2718419958890496000}  \notag \\&
+\frac{91733330193268616658399616009 \log (13)}{8698943868449587200}\bigg) e^{16} + \cdots \bigg]  , \\
\D \psi_{6L}^y =& \,\,  \frac{1}{(1-e^2)^6} \bigg(-\frac{3619517}{3780}+\frac{2086379 e^2}{1890}  
+\frac{236556689 e^4}{30240}+\frac{67767047 e^6}{20160}+\frac{777533 e^8}{6720} \bigg)    \notag \\&
+ \frac{1}{(1-e^2)^{9/2}} \bigg( \frac{12998}{15}+\frac{5251 e^2}{5}+\frac{333 e^4}{5} \bigg) . 
\end{align}

The 6.5PN term is a rational-number infinite series similar to that in the 5.5PN term and the $1/p$ expansion 
6.5PN term \eqref{eqn:6p5PN}, except for a higher power eccentricity singular factor
\begin{align}
\D \psi_{13/2}^y =& \,\,  \frac{\pi}{(1-e^2)^{13/2}} \bigg( -\frac{2620819}{2100}-\frac{5991086053 e^2}{705600}
-\frac{31762727813 e^4}{2822400}-\frac{243526100891 e^6}{81285120}  \notag \\&
-\frac{1219109013163 e^8}{16257024000}-\frac{4383296599 e^{10}}{520224768000}
-\frac{18184820155799 e^{12}}{655483207680000}    \notag \\&
+\frac{93245463971129 e^{14}}{9788549234688000}
-\frac{1478510613681403 e^{16}}{7047755448975360000}  + \cdots  \bigg) . 
\end{align}

The 7PN term reflects the split seen in \eqref{eqn:7PN}, \eqref{eqn:7PNL}, \eqref{eqn:7PNLchi}, and 
with a new 7PN $\log^2y$ term like \eqref{eqn:7PNL2}
\begin{align}
\D \psi_7^y =& \,\,  \frac{1}{(1-e^2)^7}  \bigg[ -\frac{1282190594044678657}{7041323520000}
+\frac{1316474014843 \g_E }{43659000}-\frac{3396608 \g_E ^2}{1575}
+\frac{25657561505749 \pi ^2}{2477260800}   \notag \\&
+\frac{341587582057 \pi ^4}{1006632960}
+\frac{2783260080883 \log (2)}{43659000}-\frac{5149696 \g_E  \log (2)}{1575}
-\frac{931328 \log ^2(2)}{1575}  +\frac{63488 \zeta (3)}{15}   \notag \\&
+\frac{282123979047 \log (3)}{8624000} -\frac{936036}{175} \g_E  \log (3) -\frac{936036}{175} \log (2) \log (3)  
-\frac{468018 \log ^2(3)}{175}    \notag \\&
- \frac{361328125 \log (5)}{24192}  
+ \bigg( -\frac{2138140154533483481}{2347107840000}+\frac{1128946342193 \g_E }{4851000}
-\frac{7219504 \g_E ^2}{525}    \notag \\&
+\frac{163472352568271 \pi ^2}{6606028800}  
+\frac{539506009571 \pi ^4}{335544320}+\frac{3277934193257 \log (2)}{8731800}
-\frac{79652512}{315} \g_E  \log (2)     \notag \\&
-\frac{80263696 \log ^2(2)}{175}   
-\frac{306185205247353 \log (3)}{275968000}+\frac{15912612}{175} \g_E  \log (3)
+\frac{15912612}{175} \log (2) \log (3)    \notag \\&
+\frac{7956306 \log ^2(3)}{175}    
+\frac{1017804296875 \log (5)}{1216512}+\frac{678223072849 \log (7)}{6082560}
+\frac{134944 \zeta (3)}{5}  \bigg) e^2    \notag \\&
+ \bigg( -\frac{188303307112006921}{234710784000}   
+\frac{8205409254671 \g_E }{23284800}
-\frac{1627684 \g_E ^2}{105}-\frac{2849889305071 \pi ^2}{55050240}   \notag \\&
+\frac{218483902289 \pi ^4}{268435456}+\frac{3490527652132457 \log (2)}{349272000}
+\frac{1697626904}{525} \g_E \log (2)+\frac{3404199436 \log ^2(2)}{525}    \notag \\&
+\frac{157927289197887 \log (3)}{7168000}-\frac{262324089}{400} \g_E  \log (3)  
-\frac{262324089}{400} \log (2) \log (3)-\frac{262324089 \log ^2(3)}{800}  \notag \\&
-\frac{6614382034523125 \log (5)}{1072963584}-\frac{1044921875 \g_E  \log (5)}{1008}  
-\frac{1044921875 \log (2) \log (5)}{1008}   \notag \\&
-\frac{1044921875 \log ^2(5)}{2016}  
-\frac{1107664826873969 \log (7)}{109486080}+30424 \zeta (3)  \bigg) e^4   + \cdots \bigg]  , \\
\D \psi_{7L}^y =& \,\,  \frac{1}{(1-e^2)^7}  \bigg(\frac{1316474014843}{87318000}
+\frac{1113254013233 e^2}{9702000}+\frac{7684310037263 e^4}{46569600}
+\frac{1050560559599 e^6}{55883520}  \notag \\&
-\frac{21719805083 e^8}{1411200} -\frac{52587594569 e^{10}}{88704000}+\frac{1815476071 e^{12}}{5376000}
+\frac{501487361 e^{14}}{1290240}  \notag \\&
+\frac{37402636549 e^{16}}{103219200}  + \cdots \bigg) 
+ 4 \left[\g_E +\log \left(\frac{8 (1-e^2)}{1+\sqrt{1-e^2}}\right) \right] \D \psi_{7L2}^y  
+ \D\psi_{7L}^{y,\chi} , \\
\D\psi_{7L}^{y,\chi} =&
\frac{1}{(1-e^2)^7} \bigg[ \bigg(\frac{602624 \log (2)}{225}-\frac{468018 \log (3)}{175}\bigg)
+\bigg(-\frac{155814256 \log (2)}{1575} +\frac{7956306 \log (3)}{175}\bigg) e^2  \notag \\&
+\bigg(\frac{865090292 \log (2)}{525}-\frac{262324089 \log (3)}{800}
-\frac{1044921875 \log (5)}{2016}\bigg) e^4
+\bigg(-\frac{7561620022 \log (2)}{525}  \notag \\&
-\frac{418174083 \log (3)}{200}
+\frac{11494140625 \log (5)}{1512}\bigg) e^6
+\bigg(\frac{27112112493049 \log (2)}{283500}    \notag \\&
+\frac{339134405423319 \log (3)}{7168000} 
-\frac{238205615234375 \log (5)}{4644864}-\frac{507989081563901 \log (7)}{27648000}\bigg) e^8   \notag \\&
+\bigg(-\frac{2438441188502 \log (2)}{3375}-\frac{2457755729168913 \log (3)}{7168000} 
+\frac{328215185546875 \log (5)}{1548288}   \notag \\&
+\frac{507989081563901 \log (7)}{1843200}\bigg) e^{10}
+\bigg(\frac{27531157668664681 \log (2)}{4961250}
+\frac{6481474893571248729 \log (3)}{8028160000}  \notag \\&
-\frac{2822922907568359375 \log (5)}{4682022912}
-\frac{92389498231271366573 \log (7)}{47775744000}\bigg) e^{12}  \notag \\&
+\bigg(-\frac{20936989659899360021 \log (2)}{625117500}
+\frac{171956257323847250841 \log (3)}{28098560000}  \notag \\&
+\frac{510786718173828125 \log (5)}{16387080192}
+\frac{1009210224594126146977 \log (7)}{119439360000}\bigg)e^{14}  \notag \\&
+\bigg(\frac{4619194353324708185237 \log (2)}{30005640000}
-\frac{982128291923792960826417 \log (3)}{14386462720000}  \notag \\&
+\frac{2383328610155127138671875 \log (5)}{134242960932864}
-\frac{18946304992805061334986887 \log(7)}{733835427840000}  \notag \\&
-\frac{1053921396311414387134166987 \log (11)}{251705551749120000}\bigg) e^{16} + \cdots \bigg] , \\
\D \psi_{7L2}^y =& \,\,  \frac{1}{(1-e^2)^7} \bigg( -\frac{849152}{1575}-\frac{1804876 e^2}{525}
-\frac{406921 e^4}{105}-\frac{543667 e^6}{630}-\frac{10593 e^8}{560} \bigg) .
\end{align}
Since the 7PN $\log^2y$ term is the next appearance of a leading log, its polynomial part is the same as that 
in \eqref{eqn:7PNL2}.

Like \eqref{eqn:7p5PN}, the 7.5PN term is a rational-number infinite series (times a factor of $\pi$), but 
carries an eccentricity singular factor in its $y$ expansion
\begin{align}
\D \psi_{15/2}^y =& \,\,   \frac{\pi}{(1-e^2)^{15/2}} \bigg( \frac{2782895449}{2910600}
+\frac{5474927197931 e^2}{279417600}+\frac{2889497460734527 e^4}{30177100800}
+\frac{14129642056150403 e^6}{131681894400}  \notag \\&
+\frac{264436406564274949 e^8}{10534551552000}  
+\frac{4865872318328300473  e^{10}}{3973030871040000}
+\frac{31810210018353273131 e^{12}}{93445686086860800}  \notag \\&
+\frac{19417529285629606991939 e^{14}}{104659168417284096000}
+\frac{30282407545500635935687 e^{16}}{261647921043210240000}  + \cdots \bigg) .
\end{align}

The breakdown of the 8PN term is discussed prior to the presentation of \eqref{eqn:8PN}, \eqref{eqn:8PNL}, 
\eqref{eqn:8PNLchi}, and \eqref{eqn:8PNL2}.  As mentioned there, the 8PN non-log term is too complex to recite 
in its entirety here, and its complete form is relegated to the online repositories \cite{BHPTK18,UNCGrav22}.  
We find 
\begin{align}
\D \psi_{8}^y  =&   \frac{1}{(1-e^2)^8} \bigg[ \bigg(\frac{78550205239878250993769}{28193459374080000}
-\frac{1888832198890393 \g_E }{15891876000}+\frac{177306208 \g_E^2}{11025}  \notag \\&
+\frac{569460279231731 \pi ^2}{123312537600}-\frac{623848083842333 \pi ^4}{21474836480}
-\frac{41942063811247 \log(2)}{1059458400}+\frac{2720192 \g_E  \log (2)}{11025}  \notag \\&
-\frac{520925728 \log ^2(2)}{11025}-\frac{868469344973829 \log (3)}{3139136000}
+\frac{59742279 \g_E  \log (3)}{1225}+\frac{59742279 \log (2) \log (3)}{1225}   \notag \\&
+\frac{59742279 \log^2(3)}{2450}+\frac{8570767578125 \log (5)}{96864768}
+\frac{678223072849 \log (7)}{92664000}-\frac{861696 \zeta(3)}{35}\bigg)  \notag \\&
+\bigg(\frac{237475489990768636607921}{14096729687040000}
-\frac{14786129411817037 \g_E}{7945938000}+\frac{2189268344 \g_E^2}{11025}  
-\frac{270260435561795257 \pi ^2}{739875225600}   \notag \\&
-\frac{10607716992396567 \pi^4}{85899345920} -\frac{93307044107449223 \log (2)}{7945938000}
+\frac{7769433776 \g_E  \log (2)}{2205} +\frac{71130150712 \log^2(2)}{11025}   \notag \\&
+\frac{152737472062641771 \log (3)}{20090470400} -\frac{14061034377 \g_E  \log (3)}{19600}
-\frac{14061034377 \log(2) \log (3)}{19600} -\frac{14061034377 \log ^2(3)}{39200}    \notag \\&
+\frac{129625141640625 \log (5)}{803618816} -\frac{3173828125 \g_E \log (5)}{7056}
-\frac{3173828125 \log (2) \log (5)}{7056} -\frac{3173828125 \log ^2(5)}{14112}    \notag \\&
-\frac{55100101995388051 \log(7)}{23721984000}
-\frac{11326048 \zeta (3)}{35}\bigg) e^2+\bigg(\frac{68813273776486435836697}{2349454947840000}   \notag \\&
-\frac{277281064009878713 \g_E }{42378336000}+\frac{1901024351 \g_E^2}{3675}
-\frac{3710687926688801501 \pi ^2}{2959500902400}-\frac{6040266216599961 \pi ^4}{85899345920}  \notag \\&
-\frac{38415494517661133\log (2)}{4953312000}-\frac{229436469566 \g_E  \log (2)}{4725}
-\frac{452086054463 \log ^2(2)}{4725}  -  \notag \\&
\frac{13361781707223294483 \log (3)}{200904704000}   
-\frac{84652534863 \g_E  \log (3)}{78400}-\frac{556288707663 \log (2) \log (3)}{78400}   
-\frac{84652534863 \log ^2(3)}{156800}   \notag \\&
-\frac{13614478243421620625 \log (5)}{130186248192}  
+\frac{2022790234375 \g_E  \log (5)}{84672}+\frac{2022790234375 \log (2) \log (5)}{84672}    \notag \\&
+\frac{2022790234375 \log ^2(5)}{169344}+\frac{16555297172261766443 \log (7)}{142331904000}
-\frac{31036266 \zeta (3)}{35}\bigg) e^4  + \cdots   \bigg]
, \\
\D \psi_{8L}^y =& \,\, \frac{1}{(1-e^2)^8}  \bigg( -\frac{1884153630595993}{31783752000}
-\frac{14577716249419597 e^2}{15891876000}-\frac{263990606631280313 e^4}{84756672000} \notag \\&
-\frac{317009475999791347 e^6}{101708006400}-\frac{35185381628335093 e^8}{50854003200}
-\frac{2054145845907949 e^{10}}{226017792000}-\frac{6410920059013081 e^{12}}{193729536000}  \notag \\&
-\frac{18165080552837 e^{14}}{677376000}-\frac{10104654253231 e^{16}}{481689600}
+ \cdots \bigg) 
+ 4 \left[\g_E +\log \left(\frac{8 (1-e^2)}{1+\sqrt{1-e^2}}\right) \right] \D \psi_{8L2}^y + \D \psi_{8L}^{y,\chi}  \notag \\
\D \psi_{8L}^{y,\chi} =& \frac{1}{(1-e^2)^8} \bigg[ \bigg(-\frac{70650464 \log (2)}{2205}
+\frac{59742279 \log (3)}{2450}\bigg)
+\bigg(\frac{2149292536 \log (2)}{1575}-\frac{14061034377 \log (3)}{39200}   \notag \\&
-\frac{3173828125 \log (5)}{14112}\bigg) e^2+\bigg(-\frac{837246081799 \log (2)}{33075}
-\frac{84652534863 \log (3)}{156800}+\frac{2022790234375 \log (5)}{169344}\bigg) e^4    \notag \\&
+\bigg(\frac{1794465193714 \log (2)}{6615}+\frac{77888690488341 \log (3)}{716800}
-\frac{2564871143359375 \log (5)}{16257024}   \notag \\&
-\frac{8816899947037 \log (7)}{331776}\bigg) e^6   
+\bigg(-\frac{21709548541867099 \log (2)}{7938000}-\frac{296896914153644769 \log (3)}{200704000}   \notag \\&
+\frac{142848309056640625 \log (5)}{130056192}+\frac{299112464634116791 \log (7)}{331776000}\bigg) e^8 
+ \cdots \bigg]  , \\
\D \psi_{8L2}^{y} =& \frac{1}{(1-e^2)^8} \bigg(\frac{44326552}{11025}+\frac{547317086 e^2}{11025}
+\frac{1901024351 e^4}{14700}+\frac{394272401 e^6}{4410}+\frac{2037624299  e^8}{141120}
+\frac{7800417 e^{10}}{31360} \bigg)  .
\end{align}

\end{widetext}

\subsection{Discussion}

By extending the calculation of the spin-precession invariant to a high order ($e^{16}$) in eccentricity, the 
expansions presented in the previous two subsections, when viewed by PN order, reveal eccentricity dependence 
that has parallels with that seen in the energy and angular momentum fluxes \cite{MunnEvan19a,MunnETC20,MunnEvan20a} 
and in the redshift invariant \cite{MunnEvan22a}.  The first three PN orders are closed in form and were 
found previously \cite{AkcaDempDola17}.  It is at 4PN to 9PN that our work makes new contributions.  At 4PN order 
the first appearance of a logarithmic term \cite{AkcaDempDola17} occurs.  Not surprisingly given past experience 
and the fact that the 4PN log term is a leading log \cite{MunnEvan19a}, we find it also has a closed-form expression.  

The 4PN log function then reappears in the 4PN non-log part when we regroup, or resum, that term.  This is not 
merely a trivial exercise, since the occurrence of the 4PN log term in the non-log part gathers together all of the 
dependence that is logarithmic in the eccentricity as well as the appearance of the Euler-Mascheroni constant 
$\g_E$.  This regrouping is directly analogous to what proved possible in the redshift invariant \cite{MunnEvan22a} 
and the fluxes \cite{Munn20,MunnETC20}.  Next, once terms are grouped on $\pi^2$, we see another closed-form 
function of $e$ emerge.  

The remaining transcendental numbers in $\D \psi_{4}$, which we group into a term called $\D\psi^{\chi}_4$, have a 
form that resembles the 3PN energy flux function $\chi(e)$ \cite{ArunETC08a, MunnEvan19a}.  The coefficients in 
$\chi(e)$ can be calculated to arbitrary order \cite{ForsEvanHopp16}.  We showed previously \cite{MunnEvan22a} a 
special function ($\Lambda_0(e)$) that provides complete knowledge of the analogous $\chi$-like function in the 
4PN redshift invariant.  It is possible that PN theory analysis might reveal a similar special function for 
$\D\psi^{\chi}_4$ that is based on the Newtonian quadrupole moment power spectrum \cite{MunnEvan19a}, but we have 
yet to find it.  Note that in the $y$-based PN expansion, an eccentricity function that 
appears to make the series converge as $e \rightarrow 1$ can be isolated from $\D \psi_{4L}^\chi$ by pulling out the 
function $-3 \log(1-e^2) \D \psi_{4L}$.  This procedure was first learned in working with the fluxes and 
redshift (see \cite{ForsEvanHopp16,MunnEvan19a} for more information).  What remains in the 4PN non-log term is 
(apparently) an infinite series with rational coefficients.  It is surprising that this part could not (yet) 
be manipulated into a closed form, as was possible in the fluxes and redshift invariant.  

At 5PN and 6PN we again found closed-form expressions in the log parts.  Once the log terms are known, they assist 
in allowing the 5PN and 6PN non-log parts to be segregated into important functional groupings like that found 
at 4PN, though with increasing complexity.  

7PN order marks the first appearance of a $\log^2y$, making it the next term in the (integer-order) leading-log 
sequence.  This $\D\psi_{7L2}$ term is also found to be closed in form.  The connection between $\D \psi_{7L2}$ 
and $\D \psi_{7L}$ then is seen to closely mirror the connection between $\D \psi_{4L}$ and $\D \psi_{4}$.  
This is exactly analogous to what occurs in the redshift invariant at 7PN order (see \cite{MunnEvan22a} for 
the connection in the redshift invariant and \cite{MunnEvan19a} for a detailed description of leading-logarithmic 
terms in the fluxes). 

Finally, we note that half-integer contributions begin at 5.5PN order, an infinite series with rational-number 
coefficients.  This contribution marks the first term in the half-integer leading-log sequence.  The next two 
half-integer PN terms (6.5PN and 7.5PN) are likewise series with rational coefficients.  The next half-integer 
contribution after that, at 8.5PN order (found in the repositories \cite{BHPTK18,UNCGrav22}), contains a log 
term, which is expected of the second element in the half-integer leading-log sequence.

\subsection{Comparison to numerical data on close orbits}

\begin{figure*}
\hspace{-1.5em}\includegraphics[scale=.71]{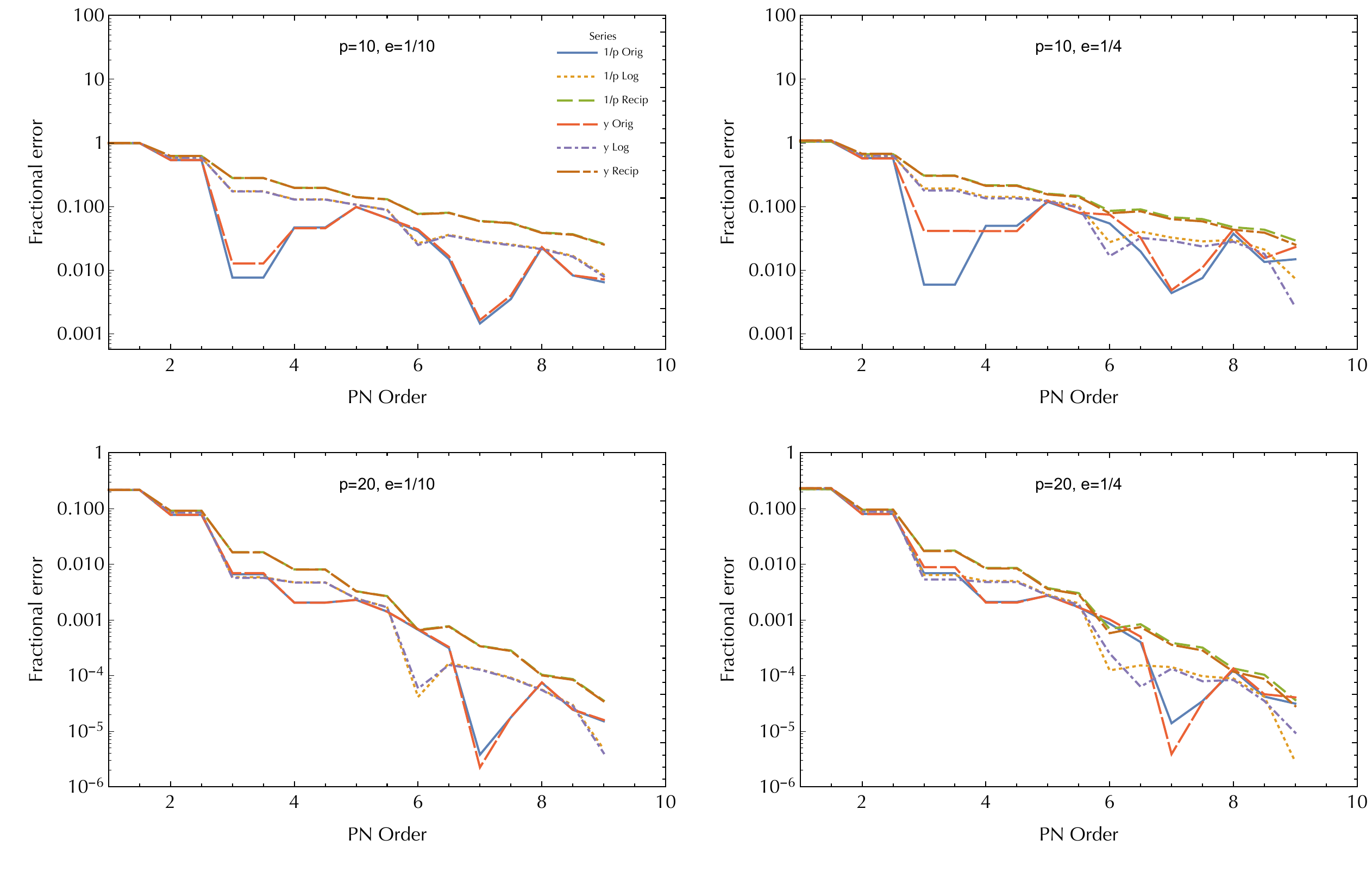}
\caption{Accuracy of the spin-precession invariant PN expansion and its resummations for several individual orbits. 
The numerical values of our redshift expansion are plotted against data from \cite{AkcaDempDola17} for the orbits 
$(p=10, e=1/10), (p=10, e=1/4), (p=20, e=1/10), (p=20, e=1/4)$.  Within each plot comparisons are made for 
both the $1/p$ and $y$ expansions, both with and without the use of logarithmic and reciprocal summations. 
Note the changes in vertical scaling in the bottom two plots. 
\label{fig:p1020ords}}
\end{figure*}

The usefulness of these high-order PN expansions in reaching into the high-speed, strong-field regime can be 
assessed by comparing their numerical evaluation to the numerical spin-precession invariant data given in the 
extensive table (Table II) of \cite{AkcaDempDola17}.  We compare both our $1/p$ and $y$ PN expansions, along 
with a few additional basic resummations applied to each.  For example, see \cite{IsoyETC13,JohnMcDa14,Munn20} on 
creating one PN series from another, like reciprocal and exponential resummation.  Our results for 
a pair of orbital sizes, $p=10$ and $p=20$, and a pair of eccentricities, $e=0.1$ and $e=0.25$, are provided 
in Fig.~\ref{fig:p1020ords}.

\begin{figure*}
\hspace{-1.5em}\includegraphics[scale=.71]{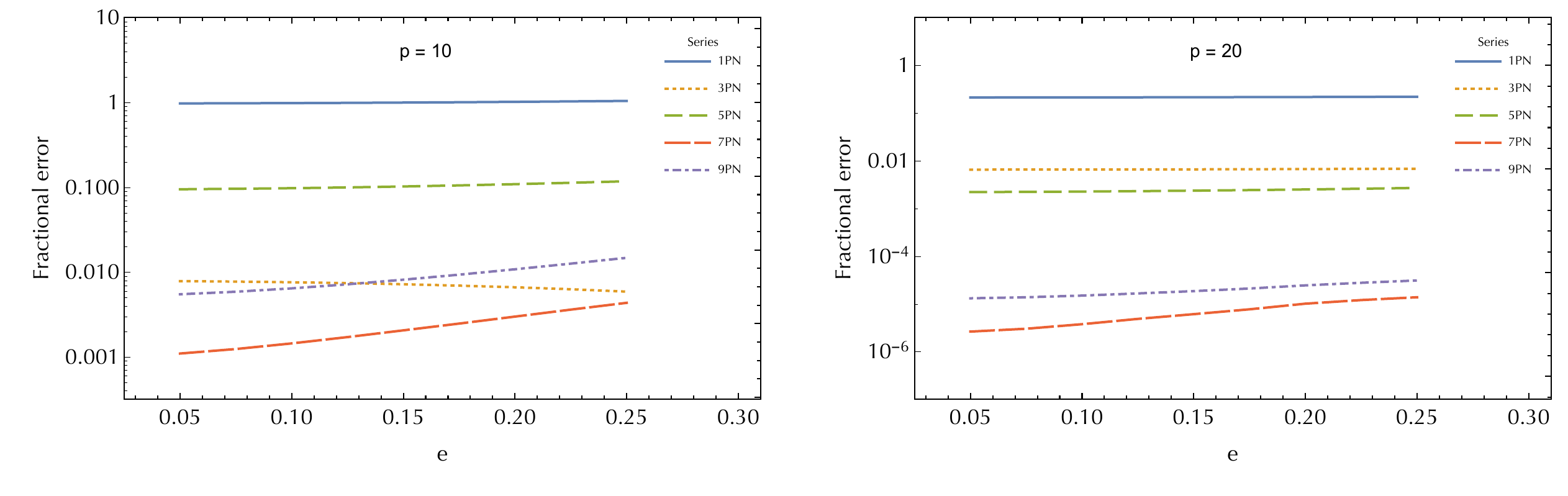}
\caption{Accuracy of the spin invariant PN expansion with increasing $e$.  The (simple) $1/p$ expansion is 
compared to numerical data for the $e$ values $0.05$ to $25$ at $0.05$ intervals (plots are made continuous for 
clarity) for both $p = 10$ and $p = 20$.  
\label{fig:eAll}}
\end{figure*}

All of the series exhibit fairly strong convergence for the case $p=20$, reaching a fractional error better than
$10^{-5}$ for both $e=0.1$ and $e=0.25$ using 9PN terms.  The dataset in \cite{AkcaDempDola17} is restricted to 
$e \le 0.25$, limiting our ability to test the expansions at higher eccentricities.  The experience with numerical 
comparisons of the redshift invariant \cite{MunnEvan22a} suggests that our series will remain viable up to 
$e \simeq 0.5$ at $p=20$.  At the closer separation of $p=10$ the convergence is markedly slower, attaining 
relative errors near $1\%$ at both $e=0.1$ and $e=0.25$.  This observation is consistent with the analysis in 
\cite{BiniDamoGera18}, who showed that the series is expected to diverge at a larger radius than the redshift 
invariant (i.e., at the separatrix, as opposed to the light ring).  Moreover, the basic resummation methods we 
have tried have not substantially improved the convergence.   

Fig.~\ref{fig:eAll} shows how the fractional errors behave versus eccentricity $e$ when using a set of series that 
have been truncated at different PN orders.   We see that knowledge gained from our calculation of the 
spin-precession invariant through $e^{16}$ provides series that converge uniformly over a range of 
eccentricity through $e=0.25$.  The curves strongly suggest that our PN series will remain accurate 
as $e \rightarrow 0.5$ or more.  The current version of our code 
could reach higher PN order but at the expense of reducing the order of the expansion in $e$, for example perhaps 
reaching 12PN and $e^{10}$.  In any event, if we consider the task of modeling EMRIs, conservative dynamical 
effects are suppressed by a factor of the mass ratio relative to the secular effect of the gravitational wave 
fluxes \cite{HindFlan08}.  Thus, our present depth of PN expansion of the spin-precession invariant is likely 
adequate for giving its contribution to EMRI dynamics.  

\section{Conclusions}
\label{sec:ConsConc}

We have presented the PN and eccentricity expansion of the spin-precession invariant $\psi$ at first order in the 
mass ratio for a point mass in bound eccentric motion about a Schwarzschild black hole.  The RWZ formalism is 
used, calculating the metric perturbation and self-force in Regge-Wheeler gauge.  The calculation is completely 
analytic, using a \textsc{Mathematica} code and drawing upon the analytic PN expansion of the MST formalism and 
the general-$l$ expansion ansatz of \cite{BiniDamo13, KavaOtteWard15}.  The construction and regularization of 
the spin-precession invariant follows methods used by \cite{AkcaDempDola17, KavaETC17, BiniDamoGera18}, as 
well as simplifications of the eccentricity dependence developed in our previous work \cite{Munn20, MunnEvan22a}.  
We have computed the spin-precession invariant to 9PN (which has been done before \cite{BiniDamoGera18, 
BiniGera19b}) but have calculated the eccentricity expansion to $e^{16}$ (with one exception, 4PN, which is 
calculated to $e^{30}$), far beyond the order $e^2$ results previously known.   

The high-order eccentricity expansions led to the discovery of five new closed-form expressions, for 
$\D \psi_{4L}$, $\D \psi_{5L}$, $\D \psi_{6L}$, $\D \psi_{7L2}$, and $\D \psi_{8L2}$.  In addition, we were able 
to use the methods developed in our past work on the gravitational wave fluxes \cite{MunnEvan19a,MunnETC20,MunnEvan20a} 
and the redshift invariant \cite{MunnEvan22a} to segregate the eccentricity dependence of many of the other PN 
terms into significant functional parts and to identify eccentricity singular factors that aid convergence in the 
$e \rightarrow 1$ limit.  The PN series results were compared to prior numerical calculations, and shown to exhibit 
fractional errors in convergence of around $10^{-5}$ for orbital separation of $p=20$ and of around $10^{-2}$ for 
$p=10$.   

The expansions of $\D\psi$ could be extended further.  The bottleneck step in the calculation is the expansion 
of the general-$l$, even-parity normalization constant $C_{lmn}^+$, which requires about 7 days on the UNC 
Longleaf cluster to reach 10PN (relative) order and $e^{20}$.  Beyond simply committing more resources or 
finding a faster cluster, intermediate expansions, which sacrifice PN order for higher order in 
eccentricity or vice versa, can be obtained immediately and may be useful.  As we described, by focusing on 
one PN order (4PN), we were able to push the eccentricity expansion to $e^{30}$.  

It will be useful now to translate these expansions to their equivalent quantities within the EOB formalism.  EOB 
waveforms have been crucial to the success of LIGO data analysis and will likely contribute to deciphering LISA 
detections.  The spin-precession invariant in first-order self-force calculations can be transcribed to yield 
portions of the EOB gyrogravitomagnetic ratio $g_{S*}(1/r;p_r;p_\vp)$, by extending a procedure described in 
\cite{KavaETC17}.  However, the process is lengthy, with each new order in $e^2$ requiring cumbersome derivations, 
and we leave that process for future work.

\acknowledgments

We thank Chris Kavanagh for helpful discussions concerning intermediate steps in the expansion procedure.  This work was supported by NSF Grant Nos.~PHY-1806447 and PHY-2110335 to the University of North Carolina--Chapel
Hill.  C.M.M.~acknowledges additional support from NASA ATP Grant 80NSSC18K1091 to MIT.

\bibliography{spin}

\end{document}